\documentclass{article}

\PassOptionsToPackage{numbers, compress}{natbib}

\usepackage{enumitem}
\usepackage{adjustbox}
\usepackage{color}
\usepackage{makecell}
\usepackage[preprint]{neurips_data_2023}
\usepackage{wrapfig}
\usepackage{hyperref}
\usepackage{amsmath,amsfonts,amssymb}
\usepackage{algorithmic}
\usepackage{array}
\usepackage[normalem]{ulem}
\usepackage{textcomp}
\usepackage{stfloats}
\usepackage{url}
\usepackage{verbatim}
\usepackage{graphicx}
\usepackage{xcolor}
\usepackage{adjustbox}
\usepackage{tabularray}
\usepackage{pifont}
\newcommand{\xmark}{\ding{55}}%
\usepackage{adjustbox}
\usepackage{array}
\usepackage{cleveref}
\usepackage{booktabs}
\usepackage{multirow}
\usepackage{color}
\usepackage{colortbl}
\usepackage{subcaption}
\usepackage{tikz}

\newcolumntype{L}{>{\centering\arraybackslash}m{1.3cm}}

\newcommand\xshlongvec[2][0]{\setlength\shlength{.5pt}%
  \stackengine{-5.6pt}{$#2$}{\smash{$\kern\shlength%
    \stackengine{7.55pt}{$\mathchar"017E$}%
      {\rule{\widthof{$#2$}}{.57pt}\kern.4pt}{O}{r}{F}{F}{L}\kern-\shlength$}}%
      {O}{c}{F}{T}{S}}

\tikzset{every picture/.style={line width=0pt}}

\usepackage[acronyms,nonumberlist,nopostdot,nomain,nogroupskip]{glossaries}

\glsresetall
\newacronym{cv}{CV}{Computer Vision}
\newacronym{dnn}{DNN}{Deep Neural Network}
\newacronym{nn}{NN}{Neural Network}
\newacronym{rl}{RL}{Reinforcement Learning}
\newacronym{drl}{DRL}{Deep Reinforcement Learning}
\newacronym{dl}{DL}{Deep Learning}
\newacronym{ml}{ML}{Machine Learning}
\newacronym{ai}{AI}{Artificial Intelligence}
\newacronym{ann}{ANN}{Artificial Neural Network}
\newacronym{cnn}{CNN}{Convolutional Neural Network}
\newacronym{fc}{FC}{Fully-connected}
\newacronym{pool}{POOL}{Pooling Layer}
\newacronym{hft}{HFT}{High-Frequency Trading}
\newacronym{svm}{SVM}{Support Vector Machine}
\newacronym{mlp}{MLP}{Multilayer Perceptron}
\newacronym{lstm}{LSTM}{Long-Short Term Memory}
\newacronym{gru}{GRU}{Gated Recurrent Unit}
\newacronym{rnn}{RNN}{Recurrent Neural Network}
\newacronym{ohlc}{OHLC}{Open High Low Close}
\newacronym{lr}{LR}{Linear Regression}
\newacronym{lob}{LOB}{Limit Order Book}
\newacronym{tabl}{TABL}{Temporal Attention-Augmented Bilinear}
\newacronym{bl}{BL}{Bilinear Layer}
\newacronym{lse}{LSE}{London Stock Exchange}
\newacronym{sota}{SOTA}{State-Of-the-Art}
\newacronym{auc}{AUC}{Area Under the Curve}
\newacronym{roc}{ROC}{Receiver Operating Characteristic}
\newacronym{tn}{TN}{True Negative}
\newacronym{tnr}{TNR}{True Negative Rate}
\newacronym{tp}{TP}{True Positive}
\newacronym{tpr}{TPR}{True Positive Rate}
\newacronym{fn}{FN}{False Negative}
\newacronym{fnr}{FNR}{False Negative Rate}
\newacronym{fp}{FP}{False Positive}
\newacronym{fpr}{FPR}{False Positive Rate}
\newacronym{seqseq}{Seq2Seq}{Sequence-to-Sequence}
\newacronym{nlp}{NLP}{Natural Language Processing}
\newacronym{mcc}{MCC}{Matthews Correlation Coefficient}
\newacronym{arima}{ARIMA}{AutoRegressive Integrated Moving Average}
\newacronym{sptp}{SPTP}{Stock Price Trend Prediction}
\newacronym{sup}{SUP}{Supplemental Material}
\newacronym{sgd}{SGD}{Stochastic Gradient Descent}

\title{LOB-Based Deep Learning Models for Stock Price Trend Prediction: A Benchmark Study}

\author{
Matteo Prata$^*$,
Giuseppe Masi$^*$, 
Leonardo Berti$^*$, 
Viviana Arrigoni$^*$, 
Andrea Coletta$^\dag$, \\ 
\textbf{Irene Cannistraci}$^*$, 
\textbf{Svitlana Vyetrenko}$^\dag$, 
\textbf{Paola Velardi}$^*$,
\textbf{Novella Bartolini}$^*$ \\
$^*$Department of Computer Science, Sapienza University of Rome, Italy. \\
$^\dag$J.P. Morgan AI Research, New York, USA. \\
\texttt{\{prata, masi.g, arrigoni, cannistraci, velardi, bartolini\}@di.uniroma1.it} \\
\texttt{\{andrea.coletta, svitlana.s.vyetrenko\}@jpmchase.com}
}

\begin{document}

\maketitle

\begin{abstract}
The recent advancements in \gls*{dl} research have notably influenced the finance sector. 
We examine the robustness and generalizability of fifteen state-of-the-art DL models focusing on \gls*{sptp} based on \gls*{lob} data.
To carry out this study, we developed LOBCAST, an open-source framework that 
incorporates data preprocessing, DL model training, evaluation and profit analysis. 
Our extensive experiments reveal that all models exhibit a significant performance drop when exposed to new data, thereby raising questions about their real-world market applicability. 
Our work serves as a benchmark, illuminating the potential and the limitations of current approaches and providing insight for innovative solutions.
\end{abstract}

\section{Introduction}
\glsresetall
Predicting stock market prices is a complex endeavour due to myriad factors, including macroeconomic conditions and investor 
sentiment~\cite{engle2013stock}. Nevertheless, professional traders and researchers usually forecast price movements by understanding key market properties, such as volatility or liquidity, and recognizing patterns to anticipate future market trends~\cite{bouchaud2018trades}. 
Effective mathematical models are essential for capturing complex market dependencies. 
The recent surge in artificial intelligence has led to significant work in using machine learning algorithms to predict future market trends~\cite{cao2022ai,jiang2021applications,sezer2020financial}. Recent \gls*{dl} models have achieved over 88\% in F1-score in predicting market trends in simulated settings using historical data~\cite{tran2021data}. However, replicating these performances in real markets is challenging, suggesting a possible \textit{simulation-to-reality} gap ~\cite{liu2022finrl,zaznov2022predicting}.

In this paper, we benchmark the most recent and promising \gls*{dl} approaches to \gls*{sptp} based on \gls*{lob} data, one of the most valuable information sources available to traders on the stock markets. Our benchmark evaluates their robustness and generalizability~\cite{pineau2021improving,gundersen2018state,baker2016reproducibility}. 
In particular, we assess the models' robustness by comparing the stated performance with our reproduced results on the same dataset FI-2010~\cite{ntakaris2018benchmark}. We also assess their generalizability by testing their performance on unseen market scenarios using LOBSTER data~\cite{lobster}.\\
We focus on novel data-driven approaches from \gls*{ml} and \gls*{dl} that analyze the market at its finest resolution, using high-frequency \gls*{lob} data. 
In this work, we formally define the \gls*{sptp} problem considering a ternary trend classification.
Our findings reveal that while best models exhibit robustness, achieving solid F1-scores on FI-2010, they show poor generalizability, as their performance significantly drops when applied to unseen LOBSTER market data.

\vspace{1cm}
The main contributions of our work are the following:
\setlist[itemize]{leftmargin=*}
\begin{itemize}
    \item We release a highly modular open-source framework called \textbf{LOBCAST}\footnote{The code is included in the supplementary material and will be publicly available upon acceptance}, to pre-process data, train, and test stock market models. Our framework employs the latest \gls*{dl} libraries to provide all researchers an easy, performing, and maintainable solution. Furthermore, to support future studies, we release two meta-learning models and a backtesting environment for profit analysis. 
    \item 
    We 
    evaluate existing LOB-based stock market trend predictors,
    showing that most of them overfit the FI-2010 dataset, with remarkably lower performance on unseen stock data.  
    \item We survey and discuss the financial performance of existing methods under different market scenarios to guide model selection in real-world applications\footnote{The details are reported in the supplementary materials for space reasons}.  
    \item We discuss the strengths and limitations of existing methodology and
    identify areas for future research toward more reliable, robust, and reproducible approaches to stock market prediction. 
\end{itemize}

\section{Related Work}\label{sec:related_work}

The increasing interest in \gls*{dl} for price trend prediction motivated several researchers to collect and analyze \gls*{sota} solutions in benchmark surveys.
The study by Jiang et al.~\cite{jiang2021applications} analyzes papers published between 2017 and 2019 that focused on stock price and market index prediction. In their literature review, the authors studied the \gls*{nn} structures and evaluation metrics used in selected papers, as well as implementation and reproducibility. This work was extended in~\cite{kumbure2022machine}, including an in-depth analysis of the data (i.e., market indices, input variables used for stock market predictions). Ozboyoglu et al. \cite{ozbayoglu2020deep} and Sezer et al. \cite{sezer2020financial} provide a comprehensive overview of the \gls*{sota} \gls*{dl} models used for financial predictions. 
The work in \cite{hu2021survey} surveys 86 papers on stock and foreign exchange price prediction. The authors review the datasets, variables, models, and performance metrics used in each surveyed article. 
In contrast to these works, in this paper, we run experiments to study the robustness and generalizability of the selected approaches.
\indent Nti et al.~\cite{nti2020systematic} conducted a systematic and critical review of 122 papers. Their study also compares the self-stated accuracy, error metrics, and software packages used in the selected papers by means of experiments. 
In contrast to this, we focus on papers that use \gls*{lob} data and \gls*{dl} algorithms for price trend predictions. 
We also evaluate the generalizability of the models by driving tests on a different dataset.
Other studies \cite{shah2022comprehensive,10050933} also analyze solutions based on sentiment analysis through \gls*{nlp} to investigate the impact of social media on the stock market, showing that this combination improves the accuracy of stock prediction models. 
In \cite{rundo2019machine}, the authors presented a comprehensive overview of traditional and \gls*{ml}-based approaches for stock market prediction and highlighted some limitations of traditional approaches, showing that \gls*{dl} models outperform them in terms of accuracy. Similar findings are reported in~\cite{mintarya2023machine}.
Lim et al.~\cite{lim2021time} discussed recent developments in hybrid \gls*{dl} models, which combine statistical and learning components for both one-step-ahead and multi-horizon time-series forecasting. Similarly, Shah et al. \cite{shah2019stock} discussed hybrid approaches in their work on the state-of-the-art algorithms commonly applied to stock market prediction. Additionally, they provided a taxonomy of computational approaches for stock market analysis and prediction.
Finally, Olorunnimbe et al. \cite{olorunnimbe2023deep} focused on exploring applications of \gls*{dl} in the stock market that involve backtesting, with a particular emphasis on research papers that meet the requirements for real-world use. They reviewed various scenarios in which \gls*{dl} has been employed in finance, with a focus on trade strategy, price prediction, portfolio management, and others.

Our work adds to this literature by providing a benchmark of recent deep learning approaches based on LOB data, evaluating their robustness and generalizability, and releasing an open-source framework for pre-processing data, training, and testing models. 

\section{Stock Price Trend Prediction}\label{sec:stock_market_pred}

The common ground that unifies the models studied in this paper is the goal of solving the \gls*{sptp} problem via \glspl*{dnn} trained on \gls*{lob} data.
\gls*{lob} data are particularly enlightening as they provide raw and granular information on stocks' trades. 
By observing the \gls*{lob} in a fixed period of time, \gls*{sptp} models return a distribution over the possible future market movements. 

\paragraph{Limit Order Book} 

A stock exchange employs a matching engine for storing and matching 
the orders issued by the trading agents. 
This is achieved by updating the so-called Limit Order Book (\gls*{lob}) data structure.
Each security (tradable asset) has a \gls*{lob}, recording all the outstanding bid and ask orders currently available on an exchange or a trading platform. The shape of the order book gives traders a simultaneous view of the market demand and supply. 

\begin{wrapfigure}[13]{r}{0.5\linewidth}
  \begin{center}
    \includegraphics[width=\linewidth]{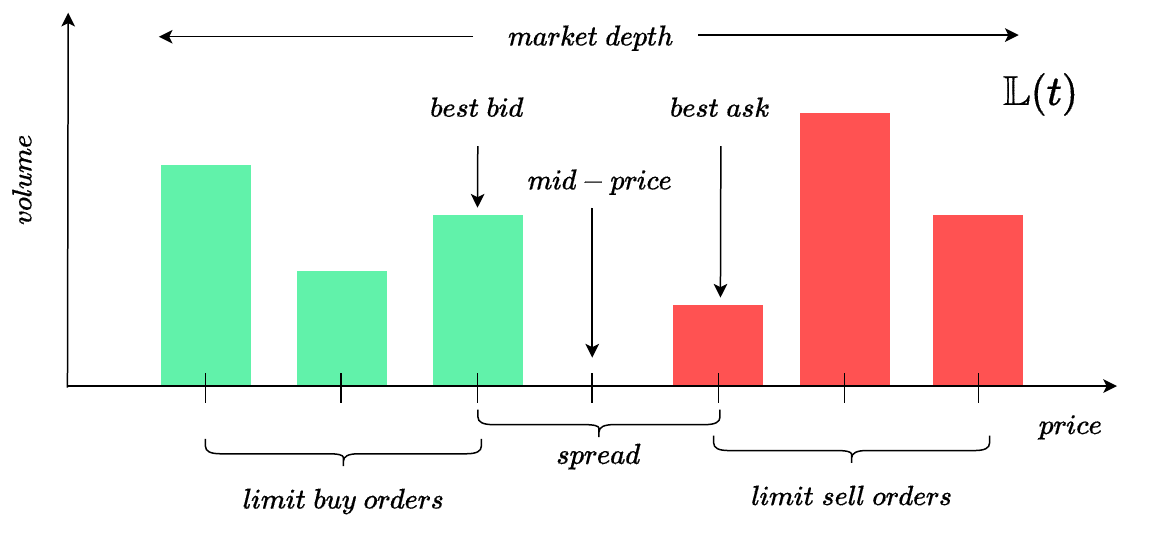}
  \end{center}
  \caption{An example of \gls*{lob}.}
    \label{fig:lob}
\end{wrapfigure}

There are three major types of orders. 
\textit{Market orders} are executed immediately at the best available price.
\textit{Limit orders}, instead, include the specification of a desired target price: a limit sell [buy] order will be executed only when it is matched to a buy [sell] order whose price is greater [lower] than or equal to the target price.  
Finally, a \textit{Cancel order} removes a previously submitted limit order.

\Cref{fig:lob} depicts an example of a \gls*{lob} snapshot, characterized by \textit{buy} orders (\textit{bid}) and \textit{sell} orders (\textit{ask}) of different prices. 
A \textit{level}, shown on the horizontal axis, represents the number of shares with the same price either on the bid or ask side. In the example of \Cref{fig:lob}, there are three bid and three ask levels. 
The \textit{best bid} is the price of the shares with the highest price on the buy side; analogously, the \textit{best ask} is the price of the shares with the lowest price on the bid side.
When the former exceeds or equals the latter, the corresponding limit ask and bid orders are executed. 
The \gls*{lob} is updated with each event (order insertion/modification/cancellation) and can be sampled at regular time intervals.

We represent the  evolution of a \gls*{lob} as a time series $\mathbb{L}$, where each $\mathbb{L}(t) \in \mathbb{R}^{4L}$ is called a LOB record, for $t=1, \ldots, N$, being $N$ the number of \gls*{lob} observations and $L$ the number of levels.
In particular, $\mathbb{L}(t) = \{P^s(t), V^{s}(t)\}_{s\in\{\texttt{ask}, \texttt{bid}\}}$, where $P^{\texttt{ask}}(t), P^\texttt{bid}(t) \in \mathbb{R}^{L}$ represent the prices of levels 1 to $L$ of the \gls*{lob}, on the \textit{ask} ($s=\texttt{ask}$) side and \textit{bid} ($s=\texttt{bid}$) side, respectively, at time $t$. 
Analogously, $V^\texttt{ask}(t), V^\texttt{bid}(t) \in \mathbb{R}^{L}$ represent the volumes. 
This means that for each $t$ and every $j\in \{1,\ldots,L\}$ on the \textit{ask} side, $V^\texttt{ask}_{j}(t)$ shares can be sold at price $P^\texttt{ask}_{j}(t)$.
The {\em mid-price} $m(t)$ of the stock at time $t$, is defined as the average value between the best bid and the best ask, 
    $m(t) = \frac{P^\texttt{ask}(t) + P^\texttt{bid}(t)}{2}$.
Mid-prices are synthetic values that are commonly used as indicators of the stock price trend. In average, if most of the executed orders are on the ask [bid] side, the mid-price increases [decreases] accordingly.

\paragraph{Trend Definition}
\label{sec:data_labelling}
We use a ternary classification for trends: \texttt{U} (``upward'') if the price trend is increasing; \texttt{D} (``downward'') for decreasing prices; and \texttt{S} (``stable'') for prices with negligible variations.
Thanks to their informativeness, mid-prices are well-suited to drive this classification. 
Nevertheless, because of the market's inherent fluctuations and shocks, they can exhibit highly volatile trends.
For this reason, using a direct comparison of consecutive mid-prices, i.e., $m(t)$ and $m(t+1)$, for stock price labelling would result in a noisy labelled dataset.
As a result, labelling strategies typically employ smoother mid-price functions instead of raw mid-prices.
Such functions consider mid-prices over arbitrarily long time intervals, called \textit{horizons}. 
Our experiments adopt the labelling proposed in~\cite{ntakaris2018benchmark} 
and repurposed in several other state-of-the-art solutions we selected for benchmarking. 
The adopted labelling strategy compares the current mid-price to the average mid-prices $a^+(k, t)$ in a future \textit{horizon} of $k$ time units, 
formally:
\begin{equation}
\footnotesize
\label{eq:avgmidprices} 
 a^+(k,t)= \frac{1}{k} \sum_{i=1}^{k} m(t+i).
\end{equation}

\normalsize
The average mid-prices are used to define a static threshold $\theta\in (0,1)$ that is used to identify an interval around the current mid-price and define the class of the trend at time $t$ as follows:
\footnotesize{
\begin{equation}
\footnotesize
\label{eq:classes}
\text{\texttt{U}:}\, a^+(k,t) > m(t)(1+\theta),\,\, \text{\texttt{D}:}\, a^+(k,t) < m(t)(1-\theta),\,\text{\texttt{S}:}\, a^+(k,t) \in [m(t)(1-\theta), m(t)(1+\theta)].
\end{equation}
}

\normalsize
With this labelling, we beat the effect of mid-price fluctuations by considering their average over a desired horizon $k$ and considering a trend to be stable when the average mid-price variations do not change significantly, thus avoiding over-fitting.
We highlight that time stamps $t$ can come either from a homogeneous or an event-based process. In our experiments, we consider an event-based process.

\paragraph{Models I/O}
\label{sec:models-io}
Given the time series of a \gls*{lob} $\mathbb{L}$ and a temporal window $T = [t-h, t]$, $h\in \mathbb{N}$, we can extract \textit{market observations} on $T$, $\mathbb{M}(T)$, by considering the sub-sequence of \gls*{lob} observations starting from time $t-h$ up to $t$.
In Section 1 of the \gls*{sup}, we give a representation of a market observation $\mathbb{M}(T) \in \mathbb{R}^{h\times 4L}$.
The market observation over the window $[t-h,t]$ is associated with the label computed through Equations~\ref{eq:avgmidprices} and~\ref{eq:classes} at time $t$. 
An \gls*{sptp} predictor takes as an input a market observation and outputs a probability distribution over the trend classes \texttt{U}, \texttt{D}, and \texttt{S}.

\section{Experiments}\label{sec:exp}

We conducted an extensive evaluation to assess the \textit{\textbf{robustness}} and \textit{\textbf{generalizability}}  of  15 \gls*{dl} models to solve the \gls*{sptp} task, as presented in \Cref{sec:stock_market_pred}. Among these, 13 were \gls*{sota} models, and 2 \gls*{dl} baseline models commonly used in the literature. More details on the models are given in \Cref{sec:models}.

In line with many other studies, we adopt the definition of robustness and generalizability introduced by J. Pineau et al. in their work~\cite{pineau2021improving}.
Robustness is evaluated by testing the proposed models on \textbf{FI-2010}, a benchmark dataset employed in all surveyed papers. In some cases, the authors of the considered works have not
provided crucial information, such as the code or the hyperparameters of their models, making reimplementation and hyperparameter search necessary. We refer to Section 5.1 in \gls*{sup} for a complete description of the hyperparameters search. To evaluate the generalizability, we created two datasets called \textbf{LOB-2021} and \textbf{LOB-2022}, extrapolated from the LOBSTER dataset~\cite{lobster}. We describe these datasets in \Cref{sec:datasets}.

Our experiments were carried out using \textbf{LOBCAST}~\cite{lobcast}, the open-source framework we make available online. The framework allows the definition of new price trend predictors based on LOB data. More details on the framework are given in \Cref{sec:lobcast}.  

\subsection{Datasets} 
\label{sec:datasets} 

\gls*{lob} data are  not often publicly available and very expensive: stock exchanges (e.g., NASDAQ) provide fine-grained data only for high fees. The high cost and low availability restrict the application and development of \gls*{dl} algorithms in the research community. 

The most widely spread public \gls*{lob} dataset is \textbf{FI-2010} which is licensed under \textit{Creative Commons Attribution 4.0 International (CC BY 4.0)} and was proposed in 2017 by Ntakaris et al.~\cite{ntakaris2018benchmark} with the objective of evaluating the performance of machine learning models on the \gls*{sptp} task. 
The dataset consists of \gls*{lob} data from five Finnish companies: Kesko Oyj, Outokumpu Oyj, Sampo, Rautaruukki, and Wärtsilä Oyj 
of the NASDAQ Nordic stock market. Data spans the time period between June 1st to June 14th, 2010, corresponding to 10 trading days (trading happens only on business days). About 4 million limit order messages are stored for 10 levels of the \gls*{lob}. 
The dataset has an event-based granularity, meaning that the time series records are not uniformly spaced in time.
\gls*{lob} observations are sampled at intervals of 10 \textit{events}, resulting in a total of 394,337 events. 
This dataset has the intrinsic limitation of being already pre-processed (filtered, normalized, and labelled) so that the original \gls*{lob} cannot be backtracked, thus hampering thorough experimentation. 
Additionally, the labelling method employed is found to be prone to instability, as demonstrated by Zhang et al. in \cite{zhang2019deeplob}. 
Moreover, the dataset is unbalanced at varying prediction horizons.   
Varying the horizon $k\in \mathcal{K} = \{1, 2, 3, 5, 10\}$, the stationary class \texttt{S} is progressively less predominant in favour of the upward and downward classes. For instance, the class composition for different values of $k$ is 
$k=1$, \texttt{U}: 18\%, \texttt{S}: 63\%, \texttt{D}:19\%; $k=5$, \texttt{U}: 32\%, \texttt{S}: 35\%, \texttt{D}:33\%; $k=10$, \texttt{U}: 37\%, \texttt{S}: 25\%, \texttt{D}:38\%.

To test the generalizability of the models in a more realistic scenario, we used data extracted from \textbf{LOBSTER}~\cite{lobster}, an online \gls*{lob} data provider for order book data, which is not available for free, as is often the case for critical applications such as health and finance~\cite{pineau2021improving}.
The data are reconstructed from NASDAQ traded stocks and are publicly available for the research community with an annual fee. 
To compare the performance of the algorithms in a wide range of scenarios, we have created a large \gls*{lob} dataset, including several stocks and time periods.
The chosen pool of stocks includes those from the top 50\% more liquid stocks of NASDAQ. 
To create a challenging evaluation scenario, we selected six stocks, namely: SoFi Technologies (SOFI), Netflix (NFLX), Cisco Systems (CSCO), Wing Stop (WING), Shoals Technologies Group (SHLS), and Landstar System (LSTR). The periods in consideration are \textit{July 2021} (2021-07-01 to 2021-07-15, 10 trading days) making up \textbf{LOB-2021}, and \textit{February 2022} (2022-02-01 to 2022-02-15, 10 trading days) making up \textbf{LOB-2022}. 
The selection of these two periods aimed to capture data from periods with different levels of market volatility. February 2022 exhibited higher volatility compared to July 2021, largely influenced by the Ukrainian crisis. This allows for an assessment of models across varying market conditions.
We describe in detail our stock selection procedure in Section 3 in \gls*{sup}.

\paragraph{Datasets for the Generalizability Study}
Due to copyright reasons, we are unable to release the LOB-2021 and LOB-2022 datasets. However, in Section 4 in \gls*{sup}, we provide detailed insights into how they are generated, ensuring transparency and replicability in future research. The approach we adopt to generate both datasets closely follows the creation process presented for FI-2010 in~\cite{ntakaris2018benchmark}. In summary, for each considered stock $s$, we construct a \textit{stock time series} of \gls*{lob} records $\mathbb{L}_s(t) \in \mathbb{R}^{4L}$, with $L=10$. To resemble the FI-2010 structure, we sample the market observation every 10 events and split records into \textit{training}, \textit{validation}, and \textit{testing} sets using a 6-2-2 days split\footnote{For the experiments on FI-2010 we followed the same data splitting procedure as the 13 \gls*{sota} papers. We split the dataset using the first 7 days for the train set and validation set (80\% / 20\%) and the last three days as the test set.}. Normalization is performed on stock time series using a $z$-score approach, and the dataset is labelled by leveraging the trend definitions described in \Cref{eq:classes}. Lastly, both LOB-2021 and LOB-2022 contain prediction labels for each one of the considered horizons $\mathcal{K}$.

\newcolumntype{P}[1]{>{\centering\arraybackslash}p{#1}}

\begin{table*}[t!]
    \centering
    \footnotesize
    \resizebox{1\textwidth}{!}{\begin{tabular}{|p{1.8cm}||P{1cm}|P{1cm}|P{1cm}|P{1cm}|P{1cm}|P{1cm}|P{1cm}|P{1cm}|P{1cm}|P{1cm}|P{1cm}|P{1cm}|P{1cm}|P{1cm}|P{1cm}|}
    \hline

&  \rotatebox{90}{\multirow{1}{*}{\begin{tabular}[c]{@{}l@{}}Tsantekidis et al.~\cite{8081663} \\ MLP (2017) \end{tabular}}} & \rotatebox{90}{\multirow{1}{*}{\begin{tabular}[c]{@{}l@{}}Tsantekidis et al.~\cite{8081663} \\ LSTM (2017) \end{tabular}}} &  \rotatebox{90}{\multirow{1}{*}{\begin{tabular}[c]{@{}l@{}}Tsantekidis et al.  \cite{tsantekidis2017forecasting} \\ CNN1 (2017) \end{tabular}}} & \rotatebox{90}{\multirow{1}{*}{\begin{tabular}[c]{@{}l@{}}Tran et al.~\cite{tran2018temporal} \\ CTABL (2018) \end{tabular}}} & \rotatebox{90}{\multirow{1}{*}{\begin{tabular}[c]{@{}l@{}}Zhang et al.~\cite{zhang2019deeplob} \\ DEEPLOB (2019) \end{tabular}}} & \rotatebox{90}{\multirow{1}{*}{\begin{tabular}[c]{@{}l@{}}Passalis et al.~\cite{passalis2019deep} \\ DAIN (2019) \end{tabular}}}  & \rotatebox{90}{\multirow{1}{*}{\begin{tabular}[c]{@{}l@{}}Tsantekidis et al. \cite{tsantekidis2020using} \\ CNNLSTM (2020) \end{tabular}}} & \rotatebox{90}{\multirow{1}{*}{\begin{tabular}[c]{@{}l@{}}Tsantekidis et al. \cite{tsantekidis2020using} \\ CNN2 (2020) \end{tabular}}} & \rotatebox{90}{\multirow{1}{*}{\begin{tabular}[c]{@{}l@{}}Wallbridge et al.  \cite{wallbridge2020transformers} \\ TRANSLOB (2020) \end{tabular}}} & \rotatebox{90}{\multirow{1}{*}{\begin{tabular}[c]{@{}l@{}}Passalis et al.  \cite{passalis2020temporal} \\ TLONBoF (2020) \end{tabular}}} & \rotatebox{90}{\multirow{1}{*}{\begin{tabular}[c]{@{}l@{}}Tran et al.  \cite{tran2021data} \\ BINCTABL (2021) \end{tabular}}} &  \rotatebox{90}{\multirow{1}{*}{\begin{tabular}[c]{@{}l@{}}Zhang et al.  \cite{zhang2021multi} \\ DEEPLOBATT (2021) \end{tabular}}} & \rotatebox{90}{\multirow{1}{*}{\begin{tabular}[c]{@{}l@{}}Guo et al.  \cite{guo2022forecasting} \\ DLA (2022)\end{tabular}}} & \rotatebox{90}{\multirow{1}{*}{\begin{tabular}[c]{@{}l@{}}Tran et al. \cite{tran2022attention} \\ ATNBoF (2022) \end{tabular}}} & \rotatebox{90}{\multirow{1}{*}{\begin{tabular}[c]{@{}l@{}}Kisiel et al.  \cite{kisiel2022axial} \\ AXIALLOB (2021) \end{tabular}}}  \\ \hline

\makecell{\textbf{temporal} \\ \textbf{shape ($h$)}} & 100 & 100 & 100 & 10 & 100 & 15 & 300 & 300 & 100 & 15 & 10 & 50 & 5 & 100 & 40 \\ \hline

\makecell{\textbf{features} \\ \textbf{shape}} & 40 & 40 & 40 & 40 & 40 & 144 & 42 & 40 & 40 & 144 & 40 & 40 & 144 & 40 & 40\\ \hline

\makecell{\textbf{code} \\ \textbf{available}} &  \xmark & \xmark & \xmark & TF & PT & PT & \xmark & \xmark & TF & PT &\xmark & TF & \xmark & PT & \xmark \\ \hline

\makecell{\textbf{n. trainable} \\ \textbf{parameters}} & $1.0 \cdot 10 ^ 6$ & $1.6 \cdot 10 ^ 4$ & $3.5 \cdot 10 ^ 4$ & $1.1 \cdot 10 ^ 4$ & $1.4 \cdot 10 ^ 5$ & $2.1 \cdot 10 ^ 6$ & $5.3 \cdot 10 ^ 4$ & $2.8 \cdot 10 ^ 5$ & $1.1 \cdot 10 ^ 5$ & $6.5 \cdot 10 ^ 5$ & $1.1 \cdot 10 ^ 4$ & $1.8 \cdot 10 ^ 5$ &  $1.2 \cdot 10 ^ 5$ & $1.3 \cdot 10 ^ 7$ & $2.0 \cdot 10 ^ 4$ \\ \hline

\makecell{\textbf{inference} \\ \textbf{time (\textit{ms})}}  & $0.08$ &  $0.21$ & $0.36$ & $0.48$ & $1.31$ & $0.15$ & $0.50$ & $0.49$ & $2.40$ & $0.43$ & $0.71$ & $1.73$ & $0.23$ & $3.90$ & $1.91$\\ \hline

    
    \end{tabular}}
    \caption{Relevant characteristics of the selected models.}
    \label{tab:selected_works}
\end{table*}

\subsection{Models} 
\label{sec:models}

We have selected 13 \gls*{sota} models based on \gls*{dl} for the \gls*{sptp} task. These models were proposed in papers published between 2017 and 2022 and utilized datasets LOB data for training and testing. In addition to the models proposed in the selected papers, we also included two classical DL algorithms, namely \gls*{mlp} and \gls*{cnn}, which were used as a benchmark in~\cite{8081663} and in~\cite{tsantekidis2020using}, respectively. 
All proposed models are based on \glspl*{dnn} and were originally trained and tested on the FI-2010 dataset.

A comprehensive summary of the benchmarked models can be found in Table~\ref{tab:selected_works}, while for additional details, we refer the reader to Section 2 in \gls*{sup}.
In Table~\ref{tab:selected_works}, the \textit{temporal shape} represents the length of the input market observation for the model. 
The \textit{features shape} refers to the number of features used by the models to infer the trend in the original papers. 
In the Table, we also indicate whether the authors released the code, and if so, whether they have used PyTorch (PT) \cite{paszke2019pytorch} or Tensorflow (TF) \cite{abadi2016tensorflow}. This is relevant  because to ensure consistency and compatibility within our proposed framework, based on PyTorch Lightning, we found it necessary to re-implement  models for which the code was not available or was only available in Tensorflow. To improve the reproducibility of the results, it is advisable for the research community to publish the code developed.

In \gls*{hft} and algorithmic trading in general, minimizing latency between model querying and order placement is of utmost importance \cite{gomber2015high}. To explore this aspect, we analyzed the inference time in milliseconds  of all models, 
based on the experiments reported in \Cref{sub:performance}.
As shown in Table~\ref{tab:selected_works}, 
DEEPLOB, DEEPLOBAT, AXIALLOB, TRANSLOB, and ATNBoF had inference times in the order of milliseconds, potentially unsuitable for HFT applications compared to other models with shorter times.
Finally, we have reported the number of trainable parameters for each model.
A noteworthy observation is that the average number of parameters is very low compared to other classical fields, such as computer vision \cite{he2016deep} and natural language processing  \cite{devlin2018bert, brown2020language}.
This leads us to conjecture that current systems 
are inadequate in effectively handling the complexity of \gls*{lob} data, as we will verify in the rest of this paper.

To explore the possibility of achieving new \gls*{sota} performance by combining the predictions of all 15 models, we have implemented two ensemble methods: 
\textit{MAJORITY}, which performs a majority voting weighted by the F1-Score of the predictions made by all the models, and \textit{METALOB}, which is trained with the predictions made by the individual models to learn the most appropriate aggregation function.
A detailed description of these ensemble methods can be found in Section 2.1 in \gls*{sup}.

\subsection{LOBCAST Framework for SPTP}
\label{sec:lobcast}
We present \textbf{LOBCAST}\footnote{https://github.com/matteoprata/LOBCAST} \cite{lobcast}, a Python-based framework developed for stock market trend forecasting using LOB data. LOBCAST is an open-source framework that enables users to test \gls*{dl} models for the \gls*{sptp} task. 
The framework provides data pre-processing functionalities, which include normalization, splitting, and labelling.
LOBCAST also offers a comprehensive training environment for \gls*{dl} models implemented in PyTorch Lightning \cite{paszke2019pytorch}. It integrates interfaces with the popular hyperparameter tuning framework WANDB \cite{wandb}, which allows users to tune and optimize model performance efficiently. The framework generates detailed reports for the trained models, including performance metrics regarding the learning task (F1, Accuracy, Recall, etc.). LOBCAST supports backtesting for profit analysis, utilizing the Backtesting.py \cite{backtesting} external library. This feature enables users to assess the profitability of their models in simulated trading scenarios.
We plan to add new features such as (i) training and testing with different \gls*{lob} representations \cite{wu2022robust, lucchese2022short}, and (ii) test on adversarial perturbations to evaluate the representations' robustness \cite{wu2021robust}.
We believe that LOBCAST, along with the advancements in DL models and the utilization of LOB data, has the potential to improve the state of the art on trend forecasting in the financial domain.

\subsection{Performance, Robustness and Generalizability} 
\label{sub:performance}
\label{sec:results}

To test robustness and generalizability, we conducted our experiments for each model using five different seeds to ensure reliable results and mitigate the impact of random initialization of network weights and training dataset shuffling. 
The training process involved training the 15 models for each seed on each of the considered prediction horizons ($\mathcal{K} = \{1, 2, 3, 5, 10\}$). More details on the setting of the experiments are provided in the \gls*{sup} Section 5.
On average over all 5 runs, the training process for all the models took approximately 155 hours for FI-2010 and 258 hours for LOB-2021/2022, utilizing a cluster comprised of 8 GPUs (1 NVIDIA GeForce RTX 2060, 2 NVIDIA GeForce RTX 3070, and 5 NVIDIA Quadro RTX 6000).
\begin{figure}[!t]
\begin{minipage}{1\textwidth}
    
\renewcommand{\arraystretch}{1.2}
\resizebox{1\columnwidth}{!}{
    \centering
    \scriptsize
     \resizebox{.9\textwidth}{!}{
     \begin{tabular}{|>{\centering\raggedleft}p{1.5cm}||>{\centering\arraybackslash}p{1.3cm}|>{\centering\arraybackslash}p{1.45cm}|>{\centering\arraybackslash}p{.5cm}|>{\centering\arraybackslash}p{1.2 cm}||>{\centering\arraybackslash}p{1.3cm}|>{\centering\arraybackslash}p{.5cm}|>{\centering\arraybackslash}p{1.2cm}||>
     {\centering\arraybackslash}p{1.3cm}|>{\centering\arraybackslash}p{.5cm}|>{\centering\arraybackslash}p{1.2cm}|}
    \hline
     & \multicolumn{4}{c||}{\textbf{FI-2010}} & \multicolumn{3}{c||}{\textbf{LOB-2021}} & \multicolumn{3}{c|}{\textbf{LOB-2022}}  \\
     \hline
    \textbf{Model} &\textbf{F1 Claim}  & \textbf{F1 LOBCAST} & \textbf{F1 Rank} & \textbf{Robustness Score (\%)} & \textbf{F1 LOBCAST} & \textbf{F1 Rank} & \textbf{General. Score (\%)}&\textbf{F1 LOBCAST} & \textbf{F1 Rank} & \textbf{General. Score (\%)} \\ 
    \hline
MLP     &$51.8 \pm 3.2$ & \textcolor{red}{$\downarrow$}\; $48.0 \pm 2.6$ &$14$   &\cellcolor[rgb]{0.3532487504805844, 0.7184159938485198, 0.3765474817377932}$91.8$        & \textcolor{green}{$\uparrow$}\;$55.5\pm 3.9$  &$14$   &\cellcolor[rgb]{0.13118031526336021, 0.6103037293348712, 0.32103037293348713}$95.0$&\textcolor{green}{$\uparrow$}\;$53.1\pm 2.5$        &$13$   &\cellcolor[rgb]{0.07597078046905037, 0.5480968858131487, 0.2887351018838908}$96.6$ \\
 \hline
LSTM    &$63.4 \pm 2.1$ & \textcolor{red}{$\downarrow$}\;$63.4 \pm 3.6$ &$7$    &\cellcolor[rgb]{0.10780469050365243, 0.5989234909650134, 0.31518646674356016}$95.5$      &\textcolor{red}{$\downarrow$}\;$56.9\pm 4.1$  &$11$   &\cellcolor[rgb]{0.6941176470588237, 0.8695886197616303, 0.44359861591695515}$85.9
$&\textcolor{red}{$\downarrow$}\;$56.1\pm 2.8$ &$9$    &\cellcolor[rgb]{0.5377931564782777, 0.8014609765474818, 0.4033064206074587}$88.8$ \\   
 \hline
CNN1    &$57.9 \pm 1.9$ & \textcolor{green}{$\uparrow$}\;$58.1 \pm 13.1$        &$10$   &\cellcolor[rgb]{0.9035755478662054, 0.9594002306805075, 0.6170703575547867}$80.9$        &\textcolor{red}{$\downarrow$}\;$57.5\pm 3.0$  &$8$    &\cellcolor[rgb]{0.06797385620915034, 0.5333333333333333, 0.28104575163398693}$97.0$&\textcolor{red}{$\downarrow$}\;$57.1\pm 2.7$        &$6$    &\cellcolor[rgb]{0.011995386389850066, 0.42998846597462514, 0.22722029988465975}$9
9.3$ \\
 \hline
CTABL   &$74.3 \pm 5.2$ &\textcolor{red}{$\downarrow$}\;$69.6 \pm 4.3$ &$5$    &\cellcolor[rgb]{0.38831218762014613, 0.7354863514033064, 0.3853133410226836}$91.3$       &\textcolor{red}{$\downarrow$}\;$59.7\pm 2.7$  &$3$    &\cellcolor[rgb]{0.9912341407151096, 0.9963091118800461, 0.7370242214532872}$78.4$&\textcolor{red}{$\downarrow$}\;$58.1\pm 3.3$  &$5$    &\cellcolor[rgb]{0.9912341407151096, 0.9963091118800461, 0.7370242214532872}$78.4$ \\  \hline
DEEPLOB &$78.9 \pm 4.4$ &\textcolor{red}{$\downarrow$}\;$71.4 \pm 5.3$ &$4$    &\cellcolor[rgb]{0.6066897347174166, 0.8316032295271051, 0.41084198385236453}$87.6$       &\textcolor{red}{$\downarrow$}\;$59.5\pm 3.0$  &$4$    &\cellcolor[rgb]{0.9959246443675509, 0.8707420222991157, 0.5386389850057669}$73.7$
&\textcolor{red}{$\downarrow$}\;$\textbf{59.5}\pm \textbf{2.9}$  &$\textbf{1}$    &\cellcolor[rgb]{0.9968473663975395, 0.9022683583237217, 0.5850826605151864}$74.7$ \\   
 \hline
DAIN    &$66.8 \pm 1.5$ &\textcolor{red}{$\downarrow$}\;$55.6 \pm 5.9$ &$11$   &\cellcolor[rgb]{0.8918877354863515, 0.954479046520569, 0.6010765090349867
}$81.4$ &\textcolor{red}{$\downarrow$}\;$55.9\pm 4.4$  &$12$   &\cellcolor[rgb]{0.9561707035755479, 0.9815455594002307, 0.689042675893887}$79.5$&\textcolor{red}{$\downarrow$}\;$54.1\pm
 2.1$   &$12$   &\cellcolor[rgb]{0.7882352941176473, 0.9101883890811228, 0.5044982698961938}$83.9$ \\
 \hline
CNNLSTM &$47.0 \pm 0.0$ &\textcolor{green}{$\uparrow$}\;$63.2 \pm 8.4$ &$8$    &\cellcolor[rgb]{0.9977700884275279, 0.930872741253364, 0.6330642060745867
}$75.7$ &\textcolor{green}{$\uparrow$}\;$57.0\pm 3.3$  &$10$   &\cellcolor[rgb]{0.5968473663975397, 0.8272971933871589, 0.40976547481737796}$87.8$&\textcolor{green}{$\uparrow$}\;$56.8\pm 2.5$ &$7$    &\cellcolor[rgb]{0.44921184159938493, 0.7627066512879662, 0.3936178392925798}$90.3$ \\ 
 \hline
CNN2    &$45.0 \pm 0.8$ &\textcolor{green}{$\uparrow$}\;$50.5 \pm 17.3$        &$12$   &\cellcolor[rgb]{0.9930026912725874, 0.7246443675509417, 0.41591695501730086}$70.5$       &\textcolor{green}{$\uparrow$}\;$55.5\pm 3.5$  &$13$   &\cellcolor[rgb]{0.662745098039216, 0.8560553633217993, 0.42329873125720896}$86.6$&\textcolor{green}{$\uparrow$}\;$55.8\pm 3.2$  &$10$   &\cellcolor[rgb]{0.5476355247981547, 0.805767012687428, 0.40438292964244527}$88.6$ \\ 
 \hline
TRANSLOB        &$\textbf{87.3} \pm \textbf{4.0}$ &\textcolor{red}{$\downarrow$}\;$59.4 \pm 2.6$ &$9$    &\cellcolor[rgb]{0.9925413302575933, 0.7015763168012302, 0.39653979238754317}$69.9$       &\textcolor{red}{$\downarrow$}\;$57.7 \pm 2.9$  &$7$    &\cellcolor[rgb]{0.9301038062283737, 0.37116493656286037, 0.23690888119953862}$64.2$&\textcolor{red}{$\downarrow$}\;$50.4\pm 6.1$        &$14$   &\cellcolor[rgb]{0.6470588235294118, 0.0, 0.14901960784313725}$56.4$ \\ 
 \hline
TLONBoF &$53.0 \pm 0.0$ &\textcolor{red}{$\downarrow$}\;$49.7 \pm 10.5$        &$13$   &\cellcolor[rgb]{0.8860438292964246, 0.9520184544405998, 0.5930795847750866}$81.5$        &\textcolor{green}{$\uparrow$}\;$57.3\pm 2.9$  &$9$    &\cellcolor[rgb]{0.015993848519800083, 0.43737024221453286, 0.2310649750096117}$\textbf{99.1}$&\textcolor{green}{$\uparrow$}\;$54.2\pm 3.1$       &$11$   &\cellcolor[rgb]{0.0, 0.40784313725490196, 0.21568627450980393}$\textbf{99.9}$ \\ 
 \hline
BINCTABL        &$80.1 \pm 6.9$ & \textcolor{green}{$\uparrow$}$\,\textbf{82.6} \pm \textbf{7.0}\,\,\,$ &$\textbf{1}$    &\cellcolor[rgb]{0.003998462129950017, 0.41522491349480967, 0.2195309496347559}$\textbf{99.7}$     &\textcolor{red}{$\downarrow$}\;$\textbf{61.2}\pm \textbf{2.7}$  &$\textbf{1}$    &\cellcolor[rgb]{0.9957708573625529, 0.8630526720492118, 0.5321799307958477}$73.5$&\textcolor{red}{$\downarrow$}\;$59.2\pm 3.3$  &$2$    &\cellcolor[rgb]{0.9946943483275663, 0.8092272202998846, 0.48696655132641287}$72.3$ \\ 
 \hline
DEEPLOBATT      &$78.8 \pm 3.1$ &\textcolor{red}{$\downarrow$}\;$67.3 \pm 9.0$ &$6$    &\cellcolor[rgb]{0.8977316416762784, 0.9569396386005383, 0.6090734332948867}$81.2$        &\textcolor{red}{$\downarrow$}\;$60.1\pm 3.0$  &$2$    &\cellcolor[rgb]{0.9977700884275279, 0.930872741253364, 0.6330642060745867}$75.7$&\textcolor{red}{$\downarrow$}\;$58.9\pm 2.8$   &$3$    &\cellcolor[rgb]{0.9966935793925413, 0.8975009611687812, 0.5770857362552863}$74.5$ \\ 
 \hline
DLA     &$78.7 \pm 0.7$ &\textcolor{red}{$\downarrow$}\;$73.4 \pm 12.1$        &$2$    &\cellcolor[rgb]{0.2597462514417532, 0.6728950403690889, 0.35317185697808534}$93.2$       &\textcolor{red}{$\downarrow$}\;$57.7\pm 3.7$  &$6$    &\cellcolor[rgb]{0.9970011534025375, 0.907035755478662, 0.5930795847750865}$74.9$&\textcolor{red}{$\downarrow$}\;$56.6\pm 2.4$   &$8$    &\cellcolor[rgb]{0.9988465974625145, 0.9642445213379469, 0.689042675893887}$76.9$ \\ 
 \hline
ATNBoF  &$67.1 \pm 5.5$ &\textcolor{red}{$\downarrow$}\;$40.9 \pm 7.7$ &$15$   &\cellcolor[rgb]{0.9637831603229527, 0.47743175701653207, 0.28581314878892733}$66.1$      &\textcolor{red}{$\downarrow$}\;$53.1\pm 3.7$  &$15$   &\cellcolor[rgb]{0.9035755478662054, 0.9594002306805075, 0.6170703575547867}$80.9$&\textcolor{red}{$\downarrow$}\;$48.0\pm 6.9$  &$15$   &\cellcolor[rgb]{0.8977316416762784, 0.9569396386005383, 0.6090734332948867}$81.2$ \\ 
 \hline
AXIALLOB        &$82.0 \pm 3.7$ &\textcolor{red}{$\downarrow$}\;$73.4 \pm 5.7$ &$3$    &\cellcolor[rgb]{0.5771626297577857, 0.8186851211072665, 0.40761245674740487}$88.2$       &\textcolor{red}{$\downarrow$}\;$59.5\pm 3.3$  &$5$    &\cellcolor[rgb]{0.9937716262975779, 0.7630911188004613, 0.44821222606689726}$71.3$&\textcolor{red}{$\downarrow$}\;$58.6\pm 2.6$ &$4$    &\cellcolor[rgb]{0.9931564782775856, 0.7323337178008458, 0.4223760092272202}$70.7$ \\ 
 \hline \hline
METALOB        & -- & $82.2 \pm 7.3$ & --    &--       &$55.9\pm 2.6$  &--    &--&$53.2\pm 1.5$ &--    &-- \\ 
 \hline
MAJORITY        & -- & $60.0 \pm 12.7$ & --    &--       &$55.5\pm 2.3$  &--    &--&$47.9\pm 2.0$ &--    &-- \\ 
 \hline
   \end{tabular}
} }

    \captionof{table}{Robustness, generalizability, and performance scores of the models. Arrows indicate whether the measured F1 of a system is higher or lower than  stated in the original paper. Colour saturation highlights systems with best (green) and worst (red) robustness and generalizability scores. }
    \label{tab:score}
\end{minipage}
\end{figure}

In~\Cref{tab:score}, we summarize the results of our experiments. As the datasets are not well balanced, we focused on F1-score; other performance metrics are reported in the \gls*{sup}. The Table compares the claimed performance of each system (column F1 Claim) with those measured in the robustness (FI-2010) and generalizability (LOB-2021 and 2022) experiments. For each dataset, we show the average performance and the standard deviation achieved by each model in all the horizons, along with its rank. 
 
To evaluate the robustness and the generalizability of the models, we compute the \textbf{robustness} and the \textbf{generalizability scores},
a value $ \leq 100$ 
that is computed as $100 - (|A| + S)$, where $A$ and $S$ are defined as follows. $A$ is the average difference between the F1-score reported in the original paper and the one that we observed in our experiments on FI-2010 for  robustness, and on LOB-2021 and LOB-2022 for  generalizability. $S$ is the standard deviation of these differences. The score penalizes models that demonstrate higher variability in their performance by subtracting the standard deviation. 
The average and standard deviation were computed over the declared horizons for each model and considering all five seeds.

\Cref{tab:score} clearly highlights the following:
\setlist[enumerate]{leftmargin=*}
\begin{enumerate}
\item Except for a few systems, there is a considerable difference between the claimed performances and those measured in both robustness and generalizability experiments. Note that while the performance gap is negative on average and considerably negative in the 
scenario of LOB-2021 and 2022, a few systems outperform the claimed results, as highlighted by the arrows in \Cref{tab:score}.
\item  All models are very sensitive to hyperparameters, in fact, they  diverged (F1-score $\leqslant 33\%$) during the hyperparameters search for about half of the runs. 
\item The ranking of  systems changes considerably if we compare the declared performances with those measured in our experiments. On the other hand, the best six systems in FI-2010 remain the same in LOB-2021 and 2022. 
\item The best-ranked systems do not consistently hold the lead in terms of robustness and generalizability - except for BINCTABL. On the contrary, some of them obtained poor generalizability scores, suggesting that they overfitted the FI-2010 dataset.
\item  Five of the best six models incorporate attention mechanisms. In particular, the best-performing model is BINCTABL, which enhances the original CTABL model by adding an Adaptive Bilinear Normalization layer, enabling joint normalization of the input time series along both temporal and feature dimensions. 
On average, BINCTABL improves the F1-score by up to $9.2\%$ compared to DLA, i.e.,  the second-best model, and up to $13\%$ compared to CTABL.
\item Regrettably, ensemble models (the last two rows in \Cref{tab:score}) do not exceed the performance of the top-performing models, which is probably due to the relatively high agreement rate among systems, as shown in Section~6 in \gls*{sup}. 
\end{enumerate}

\paragraph{Robustness on FI-2010}
As far as the robustness experiments are concerned, it is important to note that some models discussed in the literature incorporate additional market observation features for predictions. This is the case for models such as DAIN, CNNLSTM, TLOBOF, and DLA. To ensure a fair comparison among the models, we included them in our study but reduced their feature set to only the 40 raw LOB features. Due to the presence of these additional features, a strict robustness study could not be conducted for these models. However, the reduction of features did not necessarily cause a deterioration in performance: of particular interest is the case of CNNLSTM, for which the authors used stationary features derived from the LOB, stating that they were better than the raw ones. Impressively, CNNLSTM achieves the greatest average improvement of $20.9\%$ among all the models, proving that, for this model, the raw LOB features are better suited to forecast the mid-price movement than the features proposed by the original authors. 
This is also the case for DLA, which originally uses 144 input features. In fact, with  the only raw features, DLA exhibited remarkable performance, ranking second best in terms of F1-score.

\begin{figure}[t]
    \centering
    \begin{minipage}{.245\linewidth} 
    \includegraphics[width=1.15\linewidth]{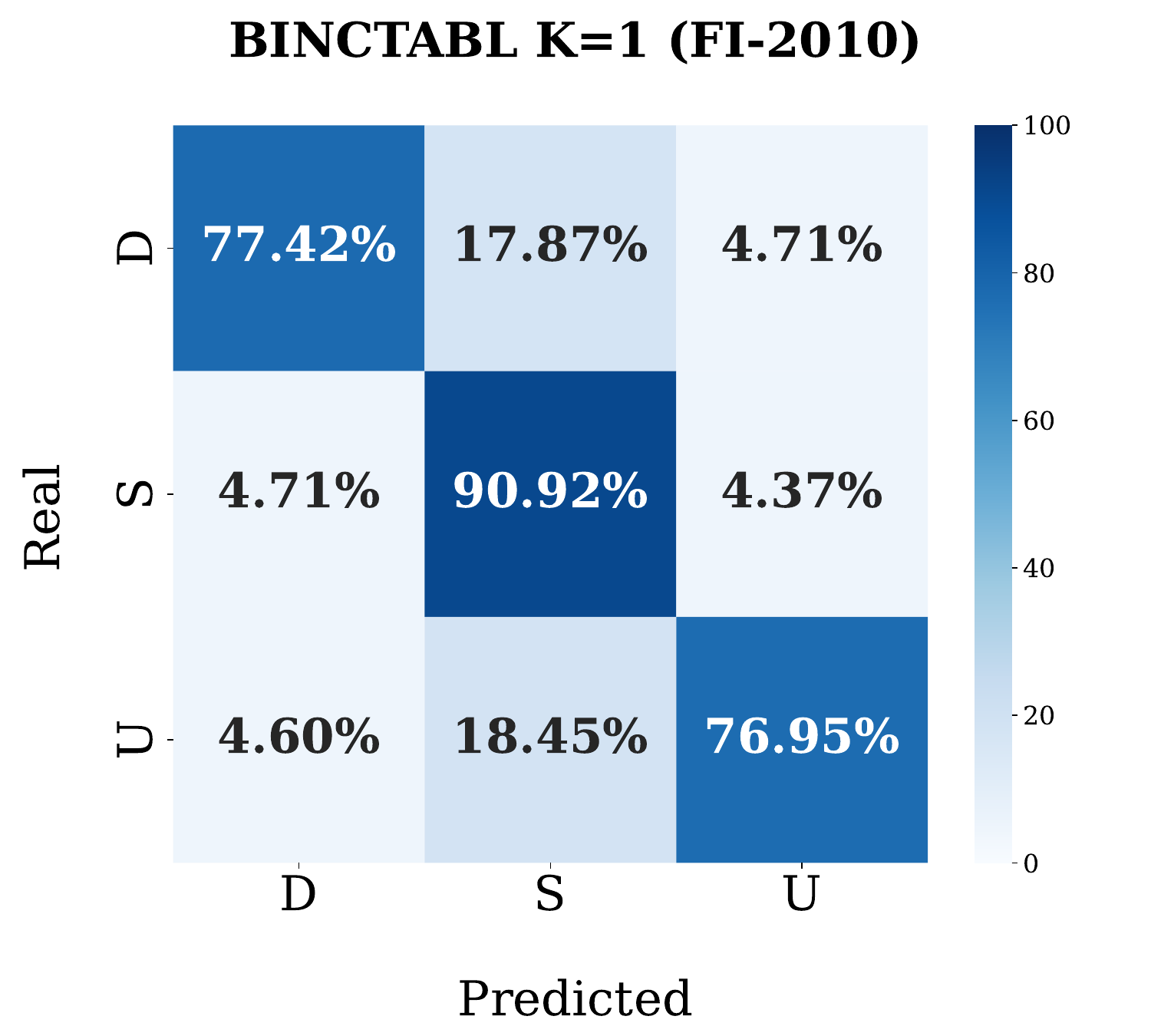}
    \label{fig:bink1f1}
    \end{minipage}
    \begin{minipage}{.245\linewidth} 
    \includegraphics[width=1.15\linewidth]{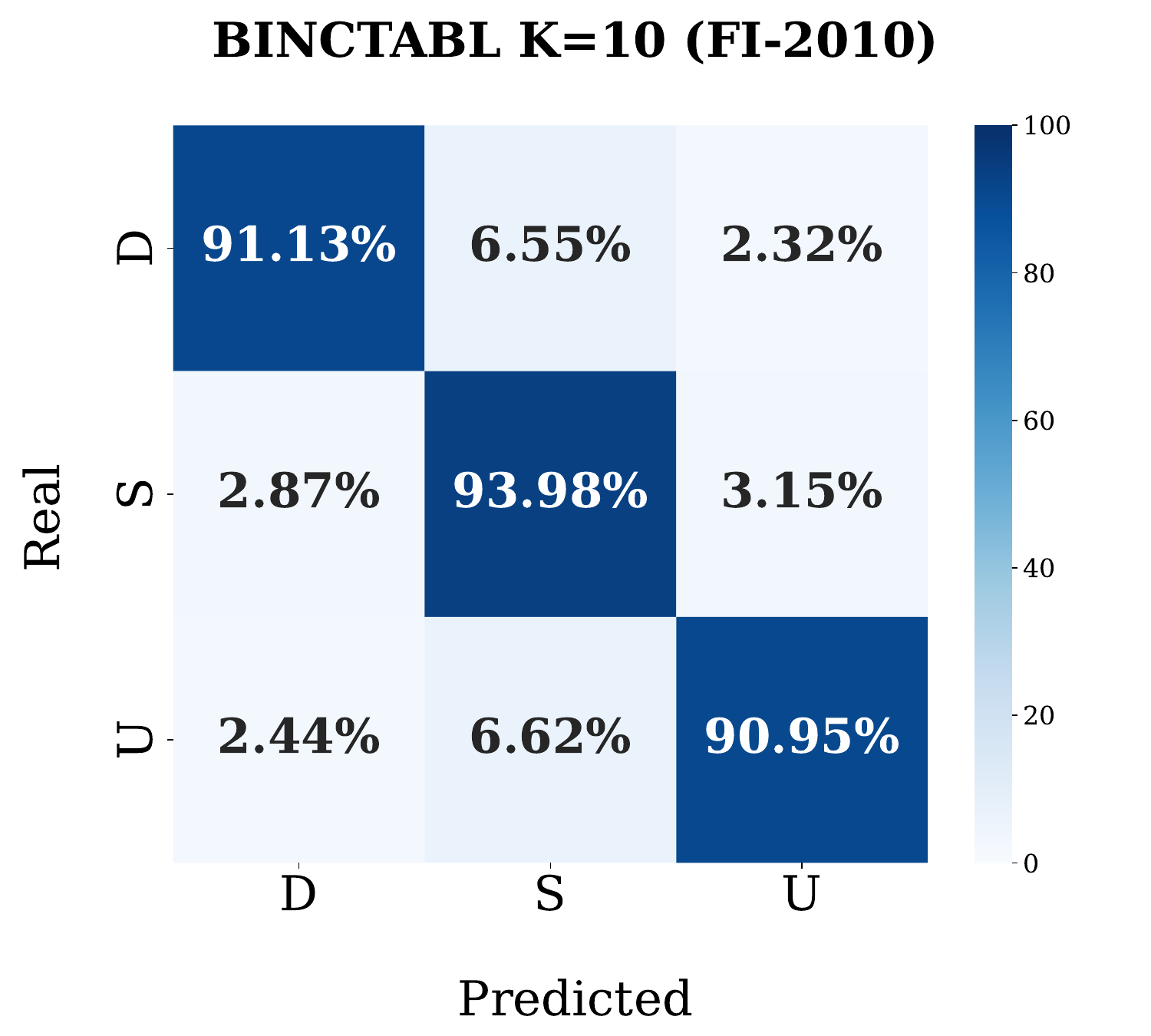}
    \label{fig:bink10f1}    
    \end{minipage}
    \begin{minipage}{.245\linewidth} 
    \includegraphics[width=1.15\linewidth]{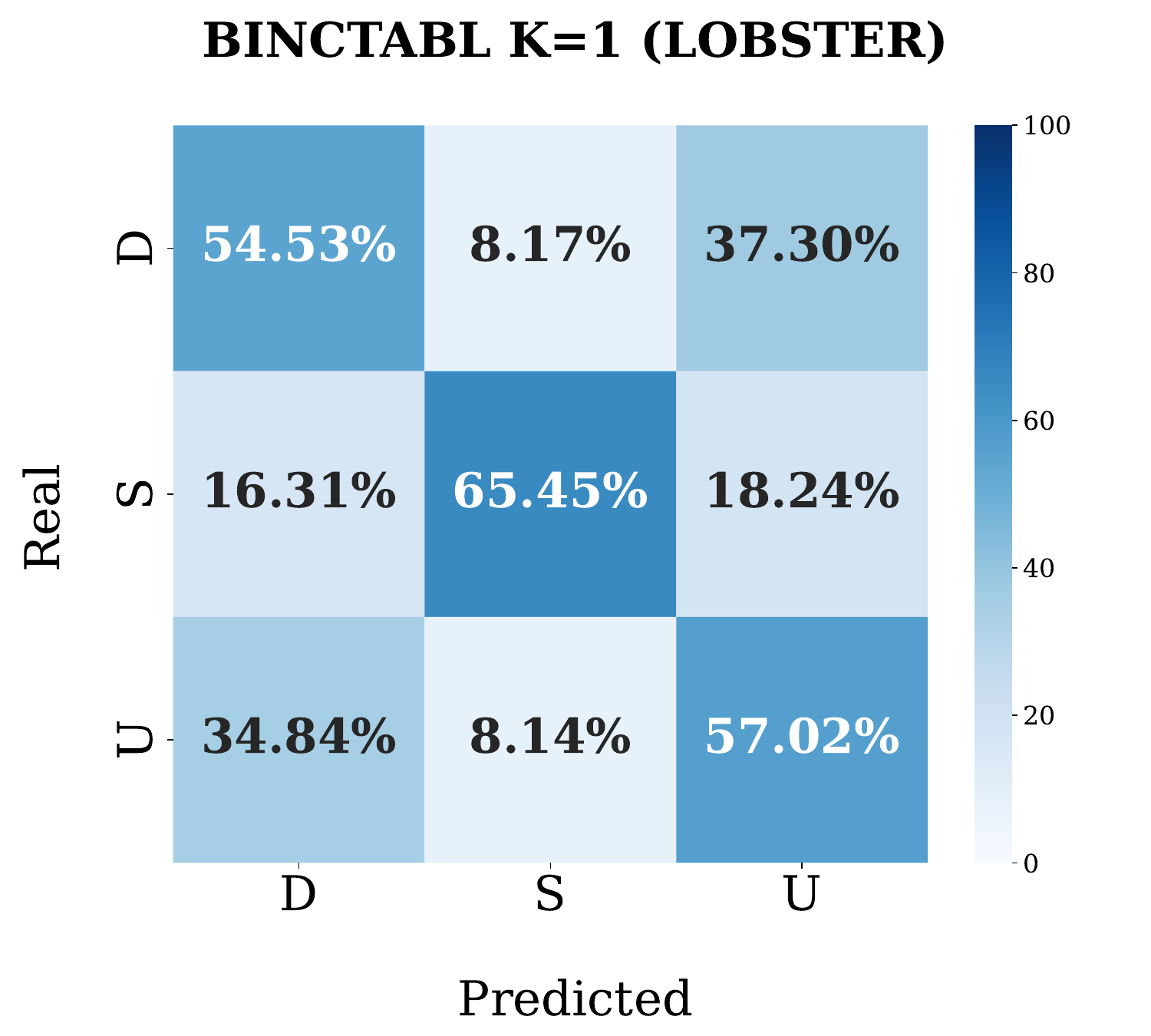}
    \label{fig:bink1lob}    
    \end{minipage}
    \begin{minipage}{.245\linewidth} 
    \includegraphics[width=1.15\linewidth]{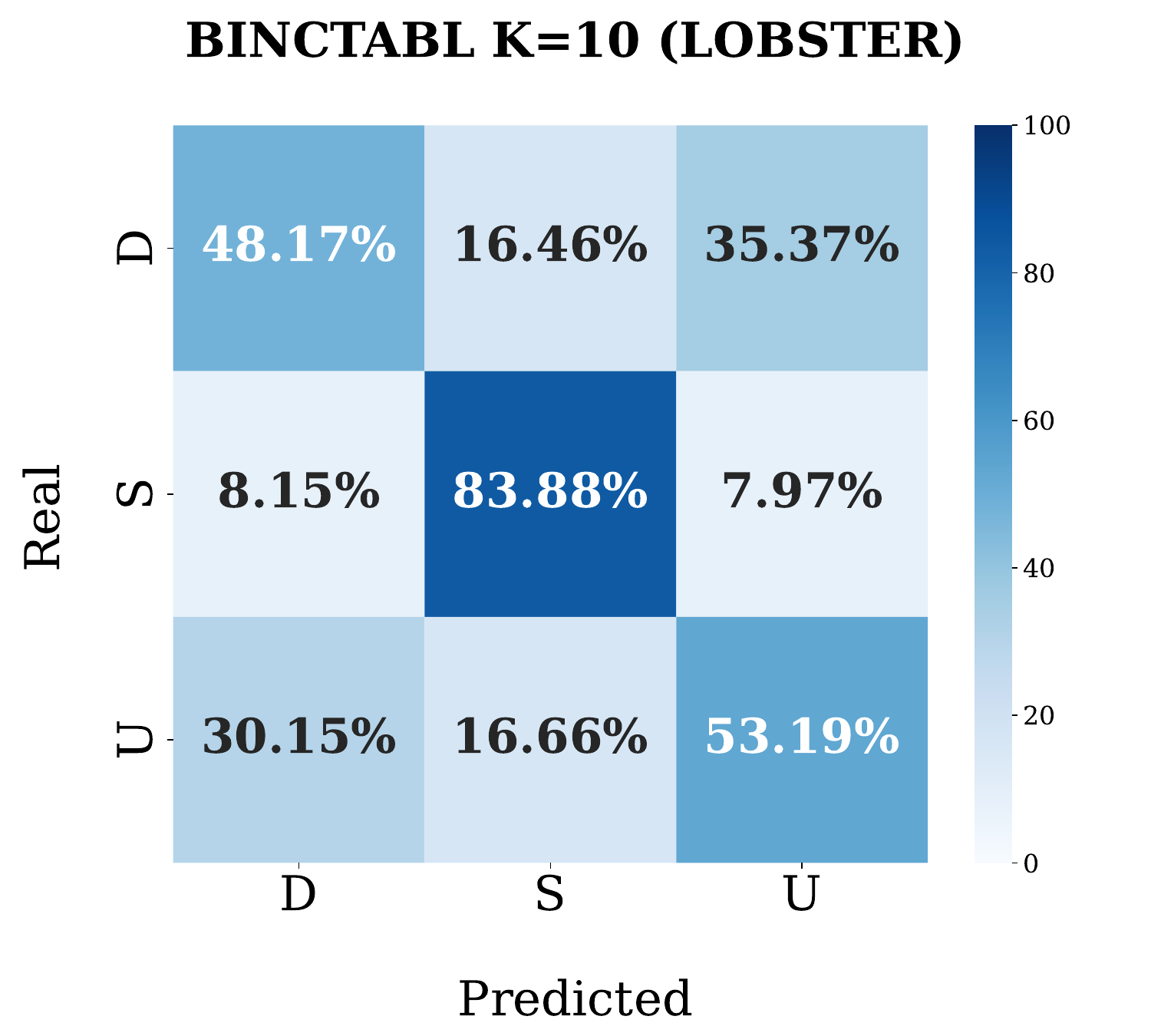}
    \label{fig:bink10lob}    
    \end{minipage}
    \captionof{figure}{Confusion matrices for BINCTABL $(k=1,10)$ on FI-2010 and LOB-2021 datasets.}
    \vspace{-.5cm}
    \label{fig:confusion_matrices}
\end{figure}

Based on these experiments (summarized in~\Cref{tab:score}), the BINCTABL model demonstrates the \textbf{highest F1-score} when averaged over the seeds and prediction horizons, achieving an average of $82.6\% \pm 7.0$. Notably, the BINCTABL model also exhibits the strongest robustness score of $99.7$, ranking as the best in terms of robustness.
For a more comprehensive analysis, \Cref{fig:confusion_matrices} provides the confusion matrices of the BINCTABL model's predictions for two horizons ($k=1$ and $k=10$). The confusion matrices demonstrate that the model is slightly biased toward the stationary class. This pattern is consistent across all the models, especially for the first three horizons, reflecting the imbalance of the dataset towards the stationary class, as specified in \Cref{sec:datasets}.\\
Remarkably, a significant number of models in our study failed to achieve the claimed performance levels. 
Two possible reasons are the lack of the original code and the missing hyperparameters declaration.
Among the models, TRANSLOB and ATNBOF exhibit the largest discrepancies, ranking as the second and first worst performers, respectively.
Notably, ATNBoF performs the poorest among all models, both in terms of robustness score and F1-score. \\
We observed that CNN1, CNN2, CNNLSTM, TLONBOF, and DLA are the most sensitive models in terms of network weight initialization and dataset shuffling, in fact, these models exhibit a standard deviation over the runs that exceeds 5 points, indicating a high degree of variability in their performance.  \\
Finally, we highlight that none of the top three models in our study utilize $h=100$ long market observations as input, despite it being a common practice in the literature~\cite{zhang2019deeplob, 8081663, tsantekidis2017forecasting, wallbridge2020transformers, tran2022attention}, meaning that they are able to achieve good results without relying on a large historical context. This suggests that the most influential and relevant dynamics impacting their predictions tend to occur within a short time frame.
In Section~6 in \gls*{sup}, we analyze in more detail the robustness results of our benchmark study when varying the horizons.

\paragraph{Generalizability on LOB-2021/2022}

When comparing the performance of models on the FI-2010 and LOB-2021/2022 datasets, we observe that  models showing high performance on the FI-2010 dataset demonstrate a deterioration in performance. Conversely, some of the models that performed poorly on the FI-2010 dataset show an improvement in performance on the LOB-2021/2022 datasets. However, the overall performance of all models on the LOB-2021/2022 dataset is still significantly lower than on the FI-2010 dataset, ranging 48-61\% in F1-score. Furthermore, we conjecture that the overall performance is worse in LOB-2022 than in LOB-2021 due to the higher stocks' volatility. 
We mention two potential factors contributing to this observed phenomenon. Firstly, the LOB-2021/2022 datasets present a higher level of complexity than the FI-2010 dataset, despite having been generated with a similar approach. Indeed, NASDAQ is a more efficient and liquid market than the Finnish one, as evidenced by the fact that LOB-2021/2022 datasets have approximately three times the size of FI-2010 in terms of events for the same period length. 
Secondly, the best-performing models may overfit the FI-2010 dataset, leading to a decrease in their performance when applied to LOB-2021/2022 datasets. 
In particular, BINCTABL experiences an average decrease of approximately $19.6\%$ in F1-Score across all horizons, resulting in a generalizability score of $73.5\%$.
For a more detailed analysis of our generalizability results, we refer to Section~6 in \gls*{sup}, 
where we also illustrate the substantial performance variation across different stocks. 
Among the tested models, CSCO stands out as yielding the highest performance. This may be attributed to the high stationarity of CSCO (balance 18-65-17\% in the train set), indicating more stable and predictable behaviour. This hypothesis is supported by the confusion matrices, which consistently show the best performance in the stationary class across all models; for reasons of space, we reported only those of BINCTABL in \Cref{fig:confusion_matrices} while for the complete study, we refer the reader to Section 7 in \gls*{sup}. 
Finally, as a final benchmark test, we conducted a trading simulation using LOB-2021. The results confirm the challenging nature of the task using the up-to-date LOB-2021 dataset, indicating that the models' profitability is far from guaranteed. For more detailed information about the simulation, please refer to Section 7 in \gls*{sup}. 

\section{Discussion and Conclusions}\label{sec:conclu}

Our findings highlight that price trend predictors based on \glspl*{dnn} using \gls*{lob} data are not consistently reliable as they often exhibit non-robust and non-generalizable performance. Our experiments demonstrate that the existing models are susceptible to hyperparameter selection, randomization, and experimental context (stocks, volatility). In addition, the selection of datasets and the experimental setup fail to capture the intricacies of the real-world scenario. This lack of generalizability makes them inadequate for practical applications in real-world settings. 

\textbf{Models}\hspace{2mm}  
Our results lead to a crucial observation: on the LOBSTER dataset, \gls*{sota} \gls*{dl} models for \gls*{lob} data exhibit low generalizability. We suggest that this phenomenon is due to two factors: the higher complexity of the LOBSTER dataset compared to the FI-2010 dataset, and the overfitting of the best-performing models to the FI-2010 dataset, which lowers their performance on the LOBSTER dataset. 
Another key finding of this study is that the top models with the highest performance on both datasets employ attention mechanisms. This suggests that the attention technique enhances the extraction of informative features and the discovery of patterns in \gls*{lob} data.
However, in general, it appears that current models cannot cope with the complexity of financial forecasting with \gls*{lob} data. Future investigations should consider state-of-the-art approaches to multivariate time series forecasting, such as \cite{lim2021temporal, drouin2022tactis, zhou2021informer}, which have not yet been adopted in the financial sector.

\textbf{Dataset}\hspace{2mm} 
Financial movements  can be influenced by geopolitical events, as political actions and decisions can significantly impact economic conditions, market sentiment, and investor confidence~\cite{engle2013stock}. These factors are not captured by \gls*{lob} data alone.
For this reason, we believe that price predictors may benefit from integrating \gls*{lob} data with additional information, for example, sentiment analysis relying on social media and press data, representing an easily accessible source of exogenous factors impacting the market \cite{ren2018forecasting}. This is particularly true for mid- and long-term price trend prediction, whereas it might not hold 
for HFT strategies~\cite{bouchaud2018trades}. 
We remark that micro and macroscopic market trends are fundamentally different, and the microscopic behaviour of the market is very much driven by HFT algorithms, making it almost exclusively dependent on financial movements rather than external factors. In this scenario, granular and raw \glspl*{lob} may suffice to provide data for price trend prediction.
Another weakness in dataset generation is the potential for training, validation, and test splits to have dissimilar distributions. This occurs due to the distinct characteristics of the historical periods covered by the stock time series. This can negatively affect the model's ability to generalize effectively and make reliable predictions on unseen data.

\textbf{Labelling}\hspace{2mm} 
As we discussed in Sections~\ref{eq:classes}, \ref{sec:datasets} and \ref{sec:results}, the choice of the threshold for class definition in Equation~\ref{eq:classes} plays a crucial role in determining the trend associated to a market observation. We believe that  current solutions present room for improvement. 
As discussed in Section~\ref{sec:datasets}, in FI-2010, the parameter $\theta$ was chosen to obtain a balanced dataset in the number of classes for the horizon $k=5$ (which is the mean value of the considered interval in the set $\mathcal{K}$). 
Thus, $\theta$ is not chosen in accordance with its financial implication but rather serves the purpose of balancing the dataset. 
We recall that the dataset is made of different stocks.
With such a labelling system, fixed $\theta$, stocks with low returns become associated with stable trends, as their behaviour is overshadowed by stocks exhibiting higher returns. 
Good practices that could be investigated are to use a weighted look-behind moving average to absorb data noise instead of mid-prices as in Equation~\ref{eq:classes} or to define a dynamically adapting $\theta$ which accounts for changing trends of a stock's mid-price. 
Moreover, the labelling approach of Equation~\ref{eq:classes}, used by all surveyed models, fails to leverage important aspects available in LOB data, including the volume, which directly influences stock volatility.
Therefore, another possible improvement is the definition and use of other insightful features that can be extrapolated from a LOB in addition to the mid-price. Such values could encapsulate  other peculiar and informative features, such as stocks' spread and volumes.

\textbf{Profit}\hspace{2mm}
In the context of stock prediction tasks, it is of utmost importance to go beyond standard statistical performance metrics such as accuracy and F1 score and incorporate trading simulations to assess the practical value of algorithms. \gls*{sptp} predictors' ultimate measure of success lies in their ability to generate profits under real market conditions. It is essential to conduct trading simulations using real simulators that go beyond testing on historical data. Recent progress has been made in the context of reactive simulators \cite{10.1145/3490354.3494411, coletta2022learning,mizuta2016brief,shi2023neural}.

We acknowledge that our study is subject to some limitations, which should be considered when interpreting our findings. 
First, we conducted a grid hyperparameter search for the models which did not specify them. 
Since hyperparameter search is not exhaustive, our chosen best hyperparameters could potentially undermine the quality of the original systems. 
Secondly, due to computational resource limitations, we could not train the benchmarked models on LOB datasets spanning longer periods, e.g., years rather than weeks. We recognize that doing so could have improved our generalizability results.

\section*{Disclaimer}
This paper was prepared for informational purposes in part by the Artificial Intelligence Research group of JPMorgan Chase \& Co. and its affiliates (``JP Morgan''), and is not a product of the Research Department of JP Morgan. JP Morgan makes no representation and warranty whatsoever and disclaims all liability, for the completeness, accuracy or reliability of the information contained herein. This document is not intended as investment research or investment advice, or a recommendation, offer or solicitation for the purchase or sale of any security, financial instrument, financial product or service, or to be used in any way for evaluating the merits of participating in any transaction, and shall not constitute a solicitation under any jurisdiction or to any person, if such solicitation under such jurisdiction or to such person would be unlawful.

\subsubsection*{Acknowledgements}
This research was funded by JPMorgan Chase AI Research Faculty award ``\textit{Understanding interdependent market dynamics: vulnerabilities and opportunities}". 
We also thank Poste Italiane for funding a Ph.D. scholarship on Financial applications of Artificial Intelligence.

\newpage
\appendix

\glsresetall

\section{Market Observation}
\begin{figure}[h!]
    \centering
    \includegraphics[width=.8\textwidth]{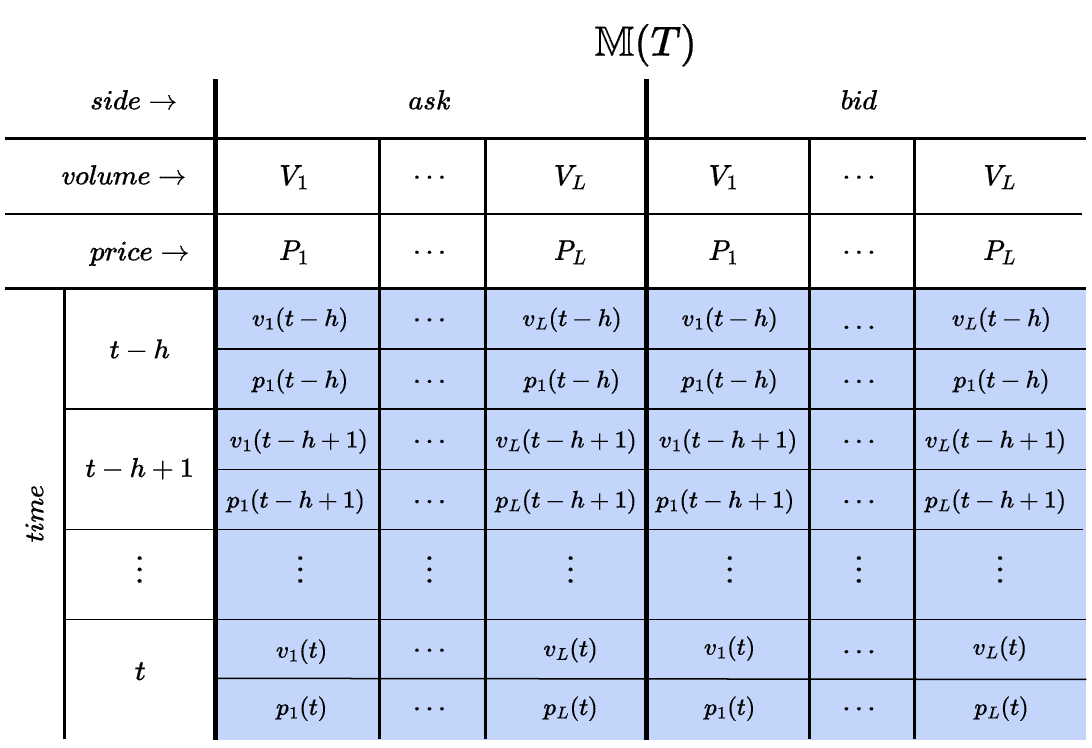}
    \caption{An example of market observation.}
    \label{fig:marketobs}
\end{figure}

We represent the evolution of a \gls*{lob} as a time series $\mathbb{L}$, where each $\mathbb{L}(t) \in \mathbb{R}^{4L}$ is called a LOB record, for $t=1, \ldots, N$, being $N$ the number of \gls*{lob} observations and $L$ the number of levels.
In particular, $\mathbb{L}(t) = \{P^s(t), V^{s}(t)\}_{s\in\{\texttt{ask}, \texttt{bid}\}}$, where $P^{\texttt{ask}}(t), P^\texttt{bid}(t) \in \mathbb{R}^{L}$ represent the prices of levels 1 to $L$ of the \gls*{lob}, on the \textit{ask} ($s=\texttt{ask}$) side and \textit{bid} ($s=\texttt{bid}$) side, respectively, at time $t$. 
Analogously, $V^\texttt{ask}(t), V^\texttt{bid}(t) \in \mathbb{R}^{L}$ represent the volumes. 
This means that for each $t$ and every $j\in \{1,\ldots,L\}$ on the \textit{ask} side, $V^\texttt{ask}_{j}(t)$ shares can be sold at price $P^\texttt{ask}_{j}(t)$.
Given the time series of a \gls*{lob} $\mathbb{L}$ and a temporal window $T = [t-h, t]$, $h\in \mathbb{N}$, we can extract \textit{market observations} on $T$, $\mathbb{M}(T)$, by considering the sub-sequence of \gls*{lob} observations starting from time $t-h$ up to $t$.
Figure~\ref{fig:marketobs} represents an observation $\mathbb{M}(T)\in \mathbb{R}^{h\times 4L}$. 
The market observation over the window $[t-h,t]$ is associated with the label computed at time $t$ through Equations 1 and 2 in the main paper. 
A \gls*{sptp} \gls*{dl} model takes as an input a market observation and outputs a probability distribution over the trend classes \texttt{U}, \texttt{D}, and \texttt{S}. 

\section{Models}
\label{sec:benchmarksdescription}
In our experiments, we consider 13 \gls*{sota} models based on \gls*{dl} for the \gls*{sptp} task. These models were published in papers between 2017 and 2022. Two additional baselines, namely \gls*{mlp} and \gls*{lstm} were also included in our analysis, in addition to two ensemble methods described in Section~\ref{sec:ensemble}. All models are trained, validated and tested with \gls*{lob} data. In the remainder of this section, we briefly describe each selected model. 

Tsantekis et al.~\cite{8081663} (2017) use a \gls*{lstm} to predict price directions considering moving averages of the mid-price over the past and the future $k$ steps. In the same year, the same authors proposed in~\cite{tsantekidis2017forecasting} a model based on a \gls*{cnn} (CNN1) for future mid-price movement predictions from large-scale high-frequency limit order data. The proposed architecture is composed of a series of convolutional and pooling layers followed by a set of fully connected layers that are used to classify the input. The parameters of the model are learnt by minimizing the categorical cross-entropy. In~\cite{tsantekidis2020using} (2020), the same research group proposed two new architectures. The first one (CNN2) uses a series of convolutional layers for capturing the temporal dynamics of time series extracted from a \gls*{lob} and for correlating temporally distant features. In the last convolutional layer, CNN2 retains the temporal ordering by flattening only the dimensions of the convolution. 
The authors then propose an architecture that merges the described CNN with an \gls*{lstm}, that we call CNNLSTM. Initially, the \gls*{cnn} is used for feature extraction for the \gls*{lob} time series. It produces a new time series of features with the same length as the original one, which is then passed to the \gls*{lstm} module for classification.

Tran et al.~\cite{tran2018temporal} in 2018 introduced a new \gls*{nn} architecture for multi-variate time series that incorporates an attention mechanism in the temporal mode. The authors call this architecture \gls*{tabl}, as it applies a bilinear transformation to the input, which consists of a set of samples at different time stamps. The \gls*{bl} is able to detect feature and time dependencies within the same sample and is augmented with a temporal attention mechanism to capture interactions between different time instances. The authors define three different network configurations, called A(\gls*{tabl}), B(\gls*{tabl}), and C(\gls*{tabl}), with 0, 1, and 2 hidden layers, respectively. In our experiments, we consider C(TABL), which outperforms the others.
In~\cite{tran2021data} (2021), the same authors extended the solutions implemented in~\cite{tran2018temporal} by integrating a data-driven normalization strategy that takes into account statistics from both temporal and feature dimensions to tackle potential problems posed by non-stationarity and multimodalities of the input series. The new model is called BINCTABL. 

Passalis (2019) et al.~\cite{passalis2019deep} introduce the DAIN (Deep Adaptive Input Normalization) three-step layer that adaptively normalizes data depending on the task at hand, instead of using some fixed statistics calculated beforehand as in traditional normalization approaches. DAIN works as follows: in the first layer, called the adaptive shifting layer, the mean of the current time series is scaled by the weight matrix of the first neural layer.
The resulting vector is passed to the adaptive scaling layer, which first computes the standard deviation of the original feature vector with respect to the shifted one, and then scales this result using the weight matrix of the scaling layer. The last layer, called the adaptive gating layer, is meant to suppress features that are not relevant by applying a sigmoid function in order to neglect features with excessive variance, which could hinder network generalization. The authors integrate DAIN in three different architectures, a \gls*{mlp} proposed in~\cite{8713851}, a \acrshort*{cnn} as in~\cite{tsantekidis2017forecasting} and \acrshort*{rnn}~\cite{cho2014learning}. In our experiments, we consider the architecture with the highest performance, namely the \gls*{mlp}.

Zhang et al.~\cite{zhang2019deeplob} (2019) propose DEEPLOB. The authors propose a smooth data labelling approach based on mid-prices to limit noise and discard small oscillations. They propose a 3-block architecture composed of standard convolutional layers, an Inception Module, and a \gls*{lstm} layer. The first two elements are used for feature extraction, whereas the  \gls*{lstm} layer captures time dependencies among the extracted features. 

Wallbridge et al.~\cite{wallbridge2020transformers} (2020) introduce TransLOB, a new \gls*{dl} architecture for mid-price trend prediction, composed of two main components: a convolutional module made up of five dilated causal convolutional layers and a transformer module, composed by two transformer encoder layers, each made up of a combination of multi-head self-attention, residual connections, normalization, and feedforward layers. Between the convolutions and the transformer module, the tensor is passed to a normalization layer and concatenated to a positional encoding. 

Passalis et al.~\cite{passalis2020temporal} (2020) propose a model for high-frequency limit order book data based on Temporal Logistic Neural Bag-of-Features formulation (TLoNBoF). Given a collection of time series, TLoNBoF extracts features with a 1-D convolutional layer to capture the temporal relationships between succeeding feature vectors. Then the features are transformed into vectors of constant length, i.e., their length must be invariant to the length of the input time series. 
To cope with this, the authors define a Temporal Logistic Neural Bag-of-Features formulation to aggregate the extracted feature vectors.
A fine-grained temporal segmentation scheme is also proposed to capture the temporal dynamics of the time series. To this end, the transformed feature vectors are segmented into three temporal regions to capture the short-term, mid-term, and long-term behaviour of the time series. 

In 2021, Zhang et al.\cite{zhang2021multi} adopt \gls*{seqseq} \cite{sutskever2014sequence, cho2014learning} and Attention \cite{luong2015effective} to recursively generate multi-horizon forecasts and build a predictor called DEEPLOBATT.  
A typical Seq2Seq model consists of an encoder that analyses the input time steps to extract meaningful features. Then, only the last hidden state from the encoder is used to make estimations, which penalizes the processing of long sequence input. To overcome this limitation, the Attention module accesses hidden states of the encoder and assigns a proper weight to each hidden state.  
Each input contains the most recent 50 updates, and each update includes information for both the ask and bid of a \gls*{lob}. Therefore, a single input has the dimension (50, 40), and each output consists of a multi-horizon prediction of all 5 points of the FI-2010 dataset. As an encoder, they adapt a previous model, namely DeepLob \cite{zhang2019deeplob}, to extract representative features from raw \gls*{lob} data while they experiment with both Seq2Seq and Attention models for the decoder.

Guo et al.~\cite{guo2022forecasting} (2022) propose a novel architecture for price trend prediction named Deep Learning Architecture (DLA). Firstly, the dataset is preprocessed and aggregated at different time windows. Once extracted, the features are given as input to the three-phase proposed architecture. The first phase uses Temporal Attention to adaptively assign attention weights to each moment of the sliding window. The processed data is passed to a stacked \gls*{gru} architecture to obtain an accurate representation of the analysed trends, which is complex and nonlinear. The \gls*{gru} architecture consists of two hidden \gls*{gru} layers to generate as output the hidden state at each time period. This is given to the second temporal attention stage, which is used to generate more accurate attention weights. 
The proposed solution is compared to several other models in the literature, including C(TABL)~\cite{tran2018temporal}, DeepLOB~\cite{zhang2019deeplob} and TLo-NBoF~\cite{passalis2020temporal}. The proposed solution achieves very high performance on the FI-2010 dataset outperforms the other models. The authors analyse the performance of their model by varying several parameters, including label thresholds and the choice of the time step.

Tran et al.~\cite{tran2022attention} extend the solution proposed in~\cite{passalis2017time}, which introduces a neural bag-of-features (N-BoF)-based method for building a \textit{codeword} that is eventually fed to a classifier. In~\cite{tran2022attention}, the neural bag-of-feature model was enhanced by incorporating a 2D-Attention (2DA) module that highlights important elements in the matrix data while discarding irrelevant ones by zeroing them out. The 2D-Attention function performs a linear interpolation between the input data matrix and input data matrix filtered by an attention mask matrix that encodes the importance of the columns of the original input. The proposed 2DA block can be applied to the features to highlight or discard the outputs of certain quantization neurons, whose results are considered equally important in the NBoF model for every input sequence (Codeword Attention). The resulting model is called ATNBoF. The 2DA function can also be applied to lend weight to salient temporal information, which is otherwise aggregated and equally contributing to the quantized features in the NBoF model (Temporal Attention). 

Kiesel et al.~\cite{kisiel2022axial} (2022) propose Axial-LOB, a model based on axial attention for price trend prediction. Unlike the naive attention mechanism, axial attention factorizes 2D attention into two 1D attention modules, one along the width (feature) axis, and a second one along the height (time) dimension. Raw values of the LOB are preprocessed and passed to the axial attention block: Each layer of the attention block is preceded and followed by a module composed of $1\times 1$ convolutions, batch normalization, and ReLu activation to adjust the number of channels in the intermediate layers of the network. For training the axial attention module, the authors use  mini-batch \gls*{sgd} by minimizing the cross-entropy loss between the predicted class probabilities and the ground truth label. The authors compare the performance of the proposed model against the solutions adopted in~\cite{tsantekidis2017forecasting,tran2018temporal,zhang2019deeplob} in terms of precision, recall, and F1 on the FI-2010 dataset. Axial-LOB proves to have improved performance with respect to these works while being simpler in terms of the number of parameters. 

\subsection{Ensamble Methods}  
\label{sec:ensemble}
To explore the possibility of achieving new \gls*{sota} performance by combining the predictions of all 15 models, we have implemented two ensemble methods: \textit{MAJORITY} and \textit{METALOB}. 

The \textit{MAJORITY} ensemble assigns the class label that appears most frequently among the predictions of the classifiers. To account for variations in the performance of individual classifiers, we incorporate a weighting scheme based on their F1 scores. This ensures that predictions from higher-performing models carry more influence in the final decision. 

The \textit{METALOB} meta-classifier is implemented as a multilayer perceptron (MLP) with two fully-connected layers. It is designed to learn how to effectively combine the outputs of the 15 DL models, which serve as the base classifiers to produce the final output. The input to the meta-classifier is a 1D tensor with a probability distribution over the trends (\textit{up, stationary, down}), for each of the models, resulting in a tensor of $3\cdot 15$ elements.  
The test set of LOB-2021/2022 is divided into three distinct subsets. We allocated 70\% of the data for training, 15\% for validating, and the remaining 15\% for testing the meta-classifier. 

By implementing these ensemble methods, our objective was to leverage the collective intelligence of ensemble models and potentially achieve performance that surpasses that of individual models. 
Unfortunately, the ensemble models did not achieve the expected level of performance, as they failed to surpass the performance of the best individual models, as discussed in the main paper.

\section{Stock Selection}
\label{sec:stock_selection}



\begin{table}
    \centering
    \scriptsize
     \begin{tblr}{|Q[.7cm,valign=m,halign=c]||Q[1.5cm,valign=m,halign=c]|Q[1.5cm,valign=m,halign=c]|Q[1.5cm,valign=m,halign=c]|Q[1cm,valign=m,halign=c]||Q[1.5cm,valign=m,halign=c]|Q[1.5cm,valign=m,halign=c]|}
    \hline 
         \textbf{Stock} & \textbf{Daily Return (\%)} & \textbf{Hourly Return (\%)} & \textbf{Market Cap.} & \textbf{P/E Ratio} & \textbf{ Train Set \\ \{ \texttt{U, S, D} \} \\ (\%), $k=5$} & \textbf{Train Set (\%), $k=5$} \\
         \hline
         SOFI & $-2.3 \pm 3.1$ & $-0.3 \pm 1.2$ & $4.26 \cdot 10^9$& $-27.84$ & $41-19-40$ & $14.8$ \\
         \hline
         NFLX & $0.6 \pm 1.7$ & $0.05 \pm 0.6$ & 1.58 $\cdot 10^{11}$ & $38.28$ & $45-5-50$  & $21.7$ \\
         \hline
         CSCO & $0.2 \pm 0.7$ & $0.02 \pm 0.4$ & $2 \cdot 10^{11}$ & $17.59$ & $18-65-17$ & $46.2$ \\
         \hline
         WING & $-0.3 \pm 3.2$ & $-0.04 \pm 0.9$ & $6.06 \cdot 10^{9}$ & $96.87$ & $44-7-49$  & $6.1$  \\
         \hline
         SHLS & $-2.4 \pm 4.9$ & $-0.3 \pm 1.9$ & $4.05 \cdot 10^{9}$ & $26.24$ & $42-14-44$ & $7.4$ \\
         \hline
         LSTR & $0.1 \pm 2.8$ & $-0.03 \pm 0.73$ & $6.16 \cdot 10^{9}$ & $16.55$ & $48-5-47$  & $3.8$ \\
         \hline
\end{tblr}
\vspace{1.0em}
\caption{Stats for the stocks (2021-07-01 -- 2021-07-15).}
    \label{tab:stocks}
\end{table}

For our generalizability study, in order to create a variegated evaluation scenario, we curated a pool of 630 stocks from NASDAQ exchange with a market capitalization that ranged from $\sim 2$ Billion to $\sim 3$ Trillion dollars. Data was gathered from NASDAQ Stock Screener~\cite{nasdaq}.
From the pool of stocks we generated 6 clusters with $t$-\textit{distributed Stochastic Neighbor Embedding} ($t$-\texttt{SNE}) to capture stocks differences in the years 2021-2023. We used the following features: daily return, hourly return, volatility, outstanding shares, P/E ratio, and market capitalization. The P/E ratio indicates the ratio between the price of a stock (P) and the company's annual earnings per share (E).
The analysis led to the identification of 6 stocks that are nearest to the cluster centroids of the generated 3-dimensional latent space. The stocks are the following: SoFi Technologies (SOFI), Netflix (NFLX), Cisco Systems (CSCO), Wing Stop (WING), Shoals Technologies Group (SHLS), and Landstar System (LSTR), making up the set that we denote by $\mathcal{S}=\{$SOFI, NFLX, CSCO, WING, SHLS, LSTR$\}$. Table~\ref{tab:stocks} captures the main features of these stocks for the period of July 2021. 
The selected stocks have very variable average daily returns, the minimum being SHLS and the maximum being NFLX. Daily and Hourly returns highlight that some stocks are more volatile than others. The market capitalization represents the total value of the outstanding common shares owned by stockholders. Stocks show different class balancing in the training set. CSCO is the stock with a major unbalance toward the stable class, whereas NFLX and LSTR are more unbalanced towards up and down classes, respectively. In Section 6, as well as in the main paper, we analyze the reasons behind the occurrence of class imbalance specific to individual stocks and discuss its impact.

\section{Datasets}

\paragraph{FI-2010} The most widely spread public \gls*{lob} dataset is \textbf{FI-2010}, which was proposed in 2017 by Ntakaris et al.~\cite{ntakaris2018benchmark} with the objective of evaluating the performance of machine learning models on the \gls*{sptp} task. 
The dataset consists of \gls*{lob} data from five Finnish companies: Kesko Oyj, Outokumpu Oyj, Sampo, Rautaruukki, and Wärtsilä Oyj (KESKOB, OUT1V, SAMPO, RTRKS, WRT1V, respectively) 
of the NASDAQ Nordic stock market. Data spans the time period between June 1st to June 14th, 2010, corresponding to 10 trading days (trading happens only on business days). About 4 million limit order messages are stored for ten levels of the \gls*{lob}. 
The dataset has an event-based granularity, meaning that the time series records are not uniformly spaced in time.
\gls*{lob} observations are sampled at intervals of 10 \textit{events} (i.e. the submission of an order that causes a LOB update), resulting in a total of 394,337 events. 

In FI-2010, the first 7 out of the 10 trading days are dedicated to the training set, while the remaining 3 days constitute the test set.
We also extracted a validation set from the training set as the last 20\% of the samples to perform hyper-parameter tuning, as in~\cite{zhang2019deeplob}.
Moreover, as FI-2010 is already normalized, we selected the dataset with $z$-score normalization for our experiments.

Finally, the dataset is already provided with the labels for each horizon $k \in \mathcal{K} = \{1, 2, 3, 5, 10\}$ by leveraging the trend definitions described in Equation~2 in the main paper.
Such a labelling scheme is very sensitive to the threshold $\theta$ regarding the resulting balancing between ``upward'', ``downward'' and ``stable'' trends.
The authors of the dataset employed a single threshold $\theta = 0.002$ for all horizons, but it balances only the case of $k = 5$. Varying the horizon $k\in \mathcal{K}$, the class imbalance occurs as shown in Table \ref{tab:stock_balancing}. Class imbalance is not addressed to guarantee a fair robustness evaluation since the considered works do not claim to have done so.

\begin{table}[t]
  \centering
  \renewcommand{\arraystretch}{1.2}
  \begin{tabular}{|c||c|c|c|}
    \hline
    Horizon $k$   & Upward (\%)  & Stationary (\%) & Downward (\%)  \\
    \hline
    1    & 18      & 63   &   19\\ \hline
    2    & 25          & 50 &  25  \\\hline
    3    & 28        & 43    &  29  \\\hline
    5    & 32        & 35    &  33  \\\hline
    10    & 37        & 25   &   38  \\
    \hline
  \end{tabular}
  \vspace{.2cm}
  \caption{Class balancing on FI-2010.}
    \label{tab:stock_balancing}
\end{table}

\begin{figure}[h!]
    \centering
    \begin{subfigure}{.49\linewidth}
        \includegraphics[width=.9\columnwidth]{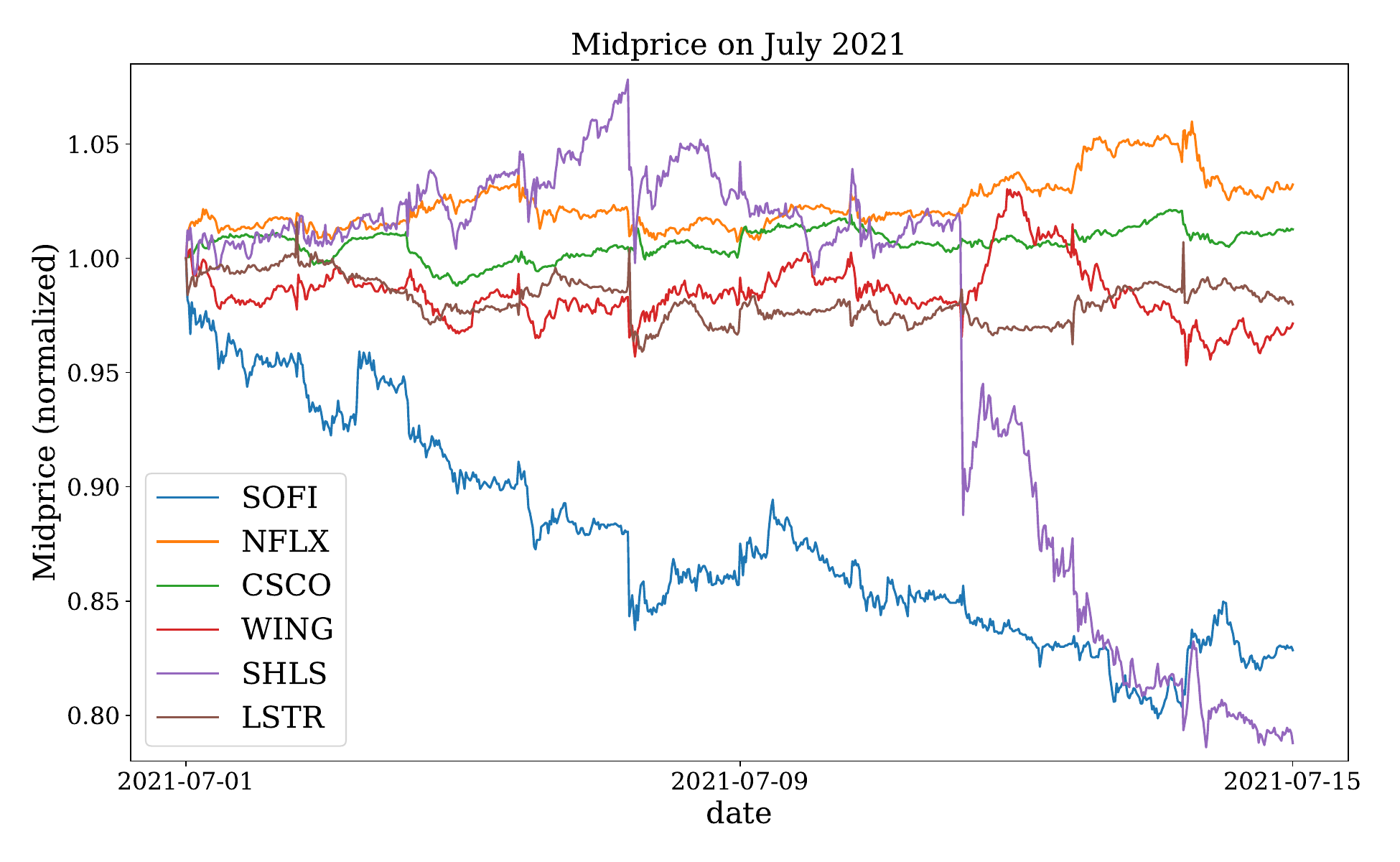}
        \caption{July 2021}
    \end{subfigure}
    \begin{subfigure}{.49\linewidth}
        \includegraphics[width=.9\columnwidth]{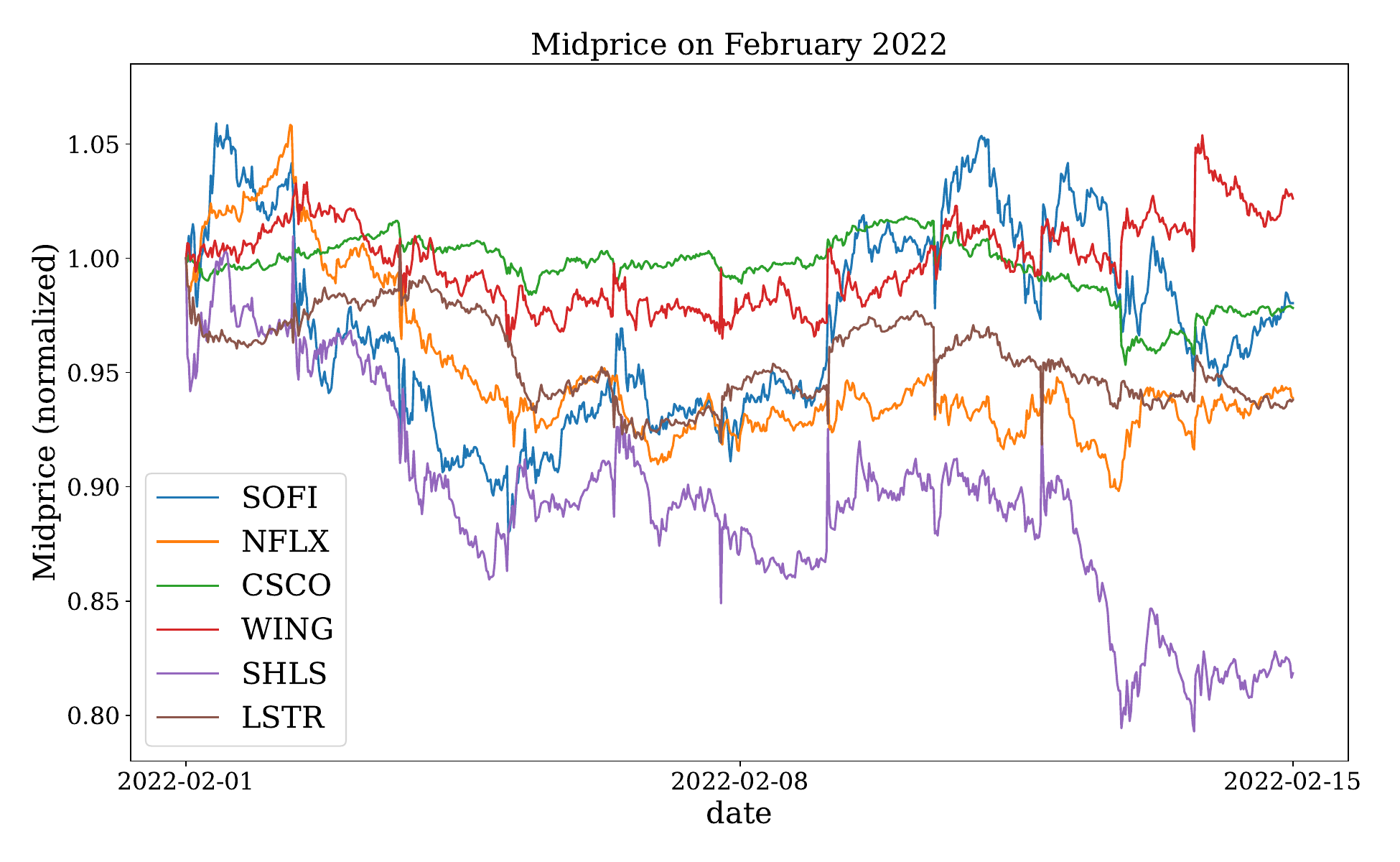}
        \caption{February 2022}
    \end{subfigure}
    \caption{Selected stocks mid-price normalized by the midprice of the first day.}
    \label{fig:stocks_trend}
\end{figure}

\paragraph{LOB-2021 and LOB-2022} To study the generalizability of the 15 models, we extracted the market observations (see Section~\ref{sec:stock_selection}) from the LOBSTER dataset in two time periods: \textit{July 2021} (2021-07-01 to 2021-07-15, ten trading days) (see \Cref{tab:stocks}) making up \textbf{LOB-2021}, and \textit{February 2022} (2022-02-01 to 2022-02-15, ten trading days) making up \textbf{LOB-2022}. 
These two periods have shown to be different in terms of volatility, as the impact of the war in Ukraine has made the market more volatile and unstable. The mid-price trends for these two periods and for the selected stocks are depicted in Figure~\ref{fig:stocks_trend}. We build the two datasets associated with these two time periods resembling the structure of the FI-2010 dataset, described in the previous section and proposed in~\cite{ntakaris2018benchmark}. 

To generate the LOB-2021/2022 datasets, we utilize the LOBSTER data, which consists of \gls*{lob} records (i.e., $\mathbb{L}(t)$ vectors) resulting from events caused by traders at the exchange. LOBSTER associates these records with the specific events that caused changes in the LOB. We isolated the following types of events: order \textit{submissions}, \textit{deletions}, and \textit{executions}, which account for almost all the events in the markets.

For each stock in the set $\mathcal{S}$ we construct a \textit{stock time series} of \gls*{lob} records $\mathbb{L}_s(t) \in \mathbb{R}^{4L}$, with $L=10, s \in \mathcal{S}$, $N_s$ being the amount of records of the stock $s$ in the considered temporal interval (e.g., (2021-07-01, 2021-07-15) for LOB-2021), $t \in [1, N_s]$. We recall that the $4\cdot 10$ features represent the prices and volumes on the buy and sell sides for the ten levels of the LOB. 
We highlight that the time series $\mathbb{L}_s$ are non-uniform in time since LOB events can occur at irregular intervals driven by traders' actions. We do not impose temporal uniformization. Instead, we sample the market observation every ten events, as for FI-2010. 
Furthermore, we do not account for liquidity beyond the 10th order level in the LOB. This approximation is necessary to ensure computational tractability while retaining the most influential levels. It is a commonly employed technique in stock market prediction models, also employed in FI-2010.

Each stock time series $\mathbb{L}_s$ is split into \textit{training}, \textit{validation}, and \textit{testing} sets using a 6-2-2 days split. Normalization is performed on stock time series using a $z$-score approach, separately normalizing the prices and volumes. The mean and standard deviation are calculated from the union of the training and validation splits for all stock time series. These statistics are then used to normalize the entire dataset, including the test splits.
The final dataset is constructed by vertical stacking (i.e., concatenating along the rows) the six training splits (i.e., one for each stock), six validation splits, and six test splits in this order.

The dataset is used to extract market observations with a sliding window approach, as explained in Section~3 of the main paper.
Labelling market observations is accomplished by leveraging the trend definitions described in Equation~2 of the main paper, mapping market observations to the corresponding trend based on a predefined prediction horizon $k\in\mathcal{K}$.
It is important to note that for each prediction horizon $\mathcal{K}$, a new dataset is generated. Consequently, LOB-2021 and LOB-2022 consist of five (i.e., $|\mathcal{K}|$) distinct datasets, each corresponding to one of the five prediction horizons.

\section{Hyperparameters Search}
For evaluating the \textit{robustness} of the surveyed models, we used the hyperparameters reported in the original papers whenever they were available. However, we encountered cases where hyperparameters were not declared at all, such as in LSTM \cite{8081663} and CNN1 \cite{tsantekidis2017forecasting}, while in other cases, including CNNLSTM \cite{tsantekidis2020using}, AXIALLOB \cite{kisiel2022axial}, ATNBOF \cite{tran2022attention} and DAIN \cite{passalis2019deep} only partial information was provided. To address these gaps, we performed a grid search exploring different values for the \textbf{batch size}, including  $\{16, 32, 64, 128, 256\}$ and the \textbf{learning rate}, including $\{0.01, 0.001, 0.0001, 0.00001\}$.

Regarding the \textit{\textit{generalizability}} experiment, we found that the majority of models using the hyperparameters from the robustness analysis performed poorly on the LOB-2021/2022 datasets. We conducted a comprehensive hyperparameter search on horizon $k=5$ (which is the most balanced) using a grid search approach for all 16 models. For this search, we maintained the same number of epochs and optimizer used in the robustness analysis, while searching for batch size and learning rate using the same domains mentioned above. 
For a complete overview of the hyperparameters utilized in our experiments, refer to Table \ref{tab:hp}.

\begin{table}[h!]
\begin{adjustbox}{width=1.0\textwidth}
\renewcommand{\arraystretch}{1.2}
\begin{tabular}{cccccc|ccccc}
\hline
                                & \multicolumn{5}{c|}{FI-2010 (Robustness)}                               & \multicolumn{5}{c}{LOB-2021/2022 (Generalizability)}                             \\ \hline
\multicolumn{1}{c|}{Model}      & Learning Rate   & Optimizer     & Batch Size   & Epochs       & Dropout & Learning Rate    & Optimizer     & Batch Size   & Epochs       & Dropout \\ \hline
\multicolumn{1}{c|}{LSTM}       & 0.001  & Adam & 32  & 100 & -       & 0.0001  & Adam & 64  & 100 & -       \\
\multicolumn{1}{c|}{MLP}        & 0.001           & Adam          & 64           & 100          & -       & 0.00001 & Adam          & 64           & 100          & -       \\
\multicolumn{1}{c|}{CNN1}       & 0.0001 & Adam & 64  & 100 & -       & 0.0001  & Adam & 32  & 100 & -       \\

\multicolumn{1}{c|}{CTABL}      & 0.01            & Adam          & 256          & 200          & -       & 0.001   & Adam          & 64  & 200          & -       \\
\multicolumn{1}{c|}{DAIN}       & 0.0001          & RMSprop       & 32  & 100 & 0.5     & 0.0001           & RMSprop       & 64  & 100 & 0.5     \\
\multicolumn{1}{c|}{DEEPLOB}    & 0.01            & Adam          & 32           & 100          & -       & 0.01             & Adam          & 32           & 100          & -       \\
\multicolumn{1}{c|}{CNNLSTM}    & 0.001  & RMSprop       & 32  & 20           & 0.1     & 0.001   & RMSprop       & 128 & 100          & 0.1     \\
\multicolumn{1}{c|}{CNN2}       & 0.001           & RMSprop       & 32           & 100          & -       & 0.001            & RMSprop       & 128 & 100          & -       \\
\multicolumn{1}{c|}{TRANSLOB}   & 0.0001          & Adam          & 32           & 150          & -       & 0.001            & Adam          & 128 & 100          & -       \\
\multicolumn{1}{c|}{TLONBoF}    & 0.0001          & Adam          & 128          & 100          & -       & 0.00001          & Adam          & 32  & 100          & -       \\
\multicolumn{1}{c|}{BINCTABL}   & 0.001           & Adam          & 128          & 200          & -       & 0.001   & Adam          & 32  & 200          & -       \\
\multicolumn{1}{c|}{DEEPLOBATT} & 0.001           & Adam          & 32           & 100          & -       & 0.0001           & Adam          & 128 & 100          & -       \\
\multicolumn{1}{c|}{AXIALLOB}   & 0.01            & SGD           & 64           & 50           & -       & 0.01    & SGD           & 64           & 50           & -       \\
\multicolumn{1}{c|}{ATNBoF}     & 0.001           & Adam          & 128 & 80           & 0.2     & 0.00001          & Adam          & 32  & 80           & 0.2     \\
\multicolumn{1}{c|}{DLA}        & 0.01            & Adam          & 256          & 100          & -       & 0.001            & Adam          & 64  & 100          & -       \\ \hline
\multicolumn{1}{c|}{METALOB}        & 0.0001            & SGD          & 64          & 100          & -       & 0.0001            & SGD          & 64  & 100          & -       \\ \hline
\end{tabular}
\end{adjustbox}
\medskip
\caption{Hyperparameters adopted in our experiments.}
\label{tab:hp}
\end{table}

\section{Additional Experimental Results}

\begin{figure}
        \centering
        \begin{subfigure}[b]{.49\columnwidth}
            \includegraphics[width=.95\columnwidth]{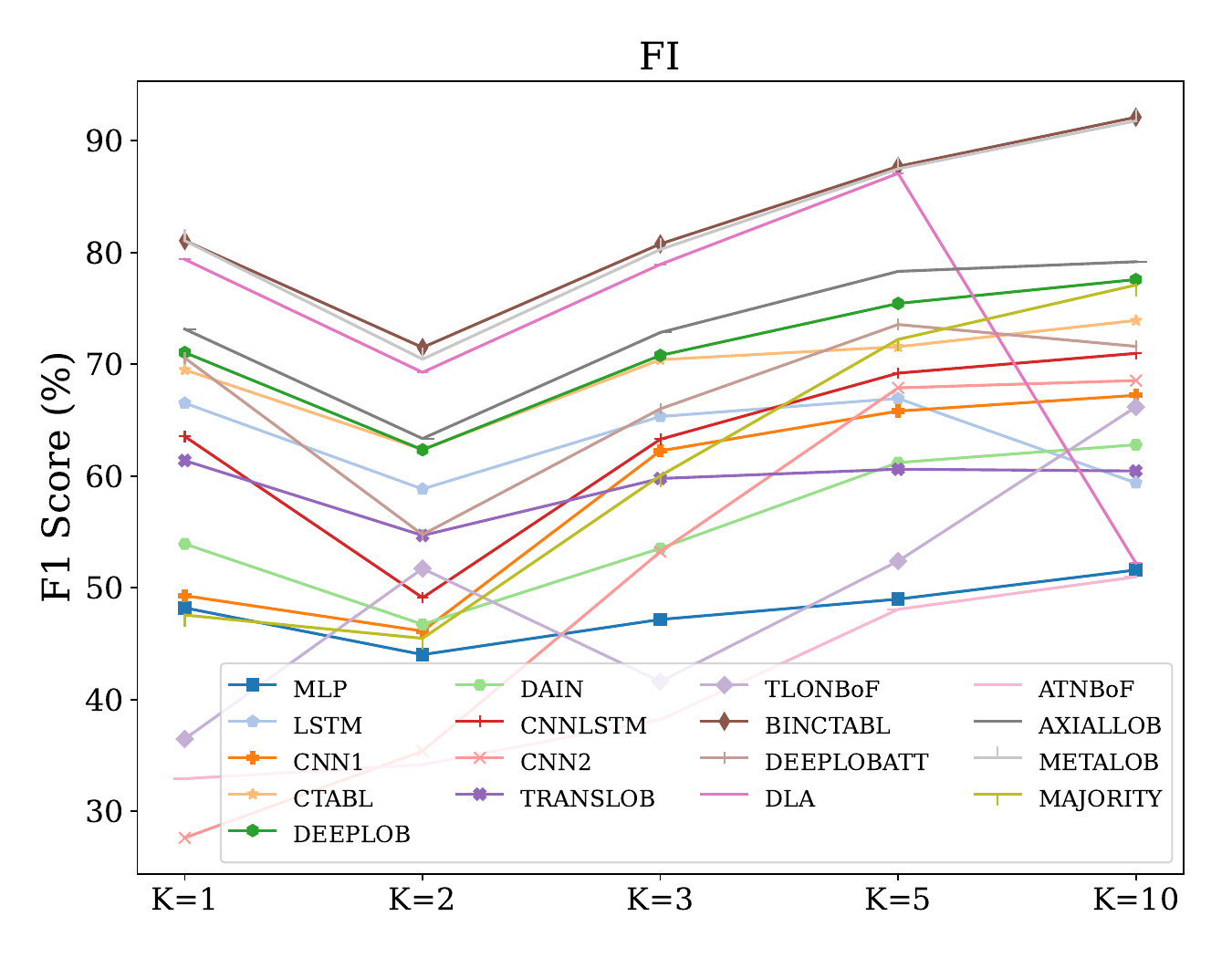}
            \caption{F1-Score}
        \end{subfigure}
        \begin{subfigure}[b]{.49\columnwidth}
            \includegraphics[width=.95\columnwidth]{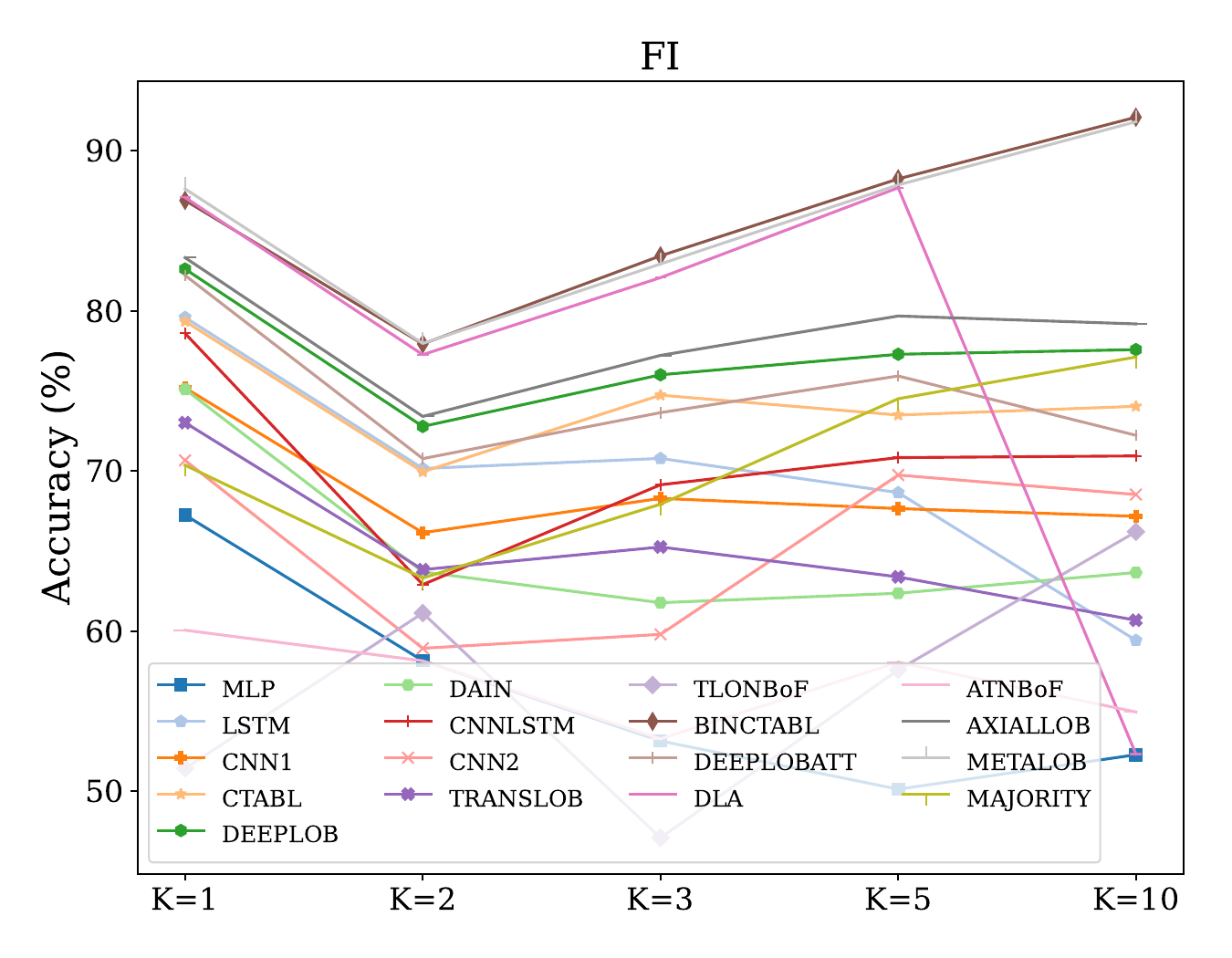}
            \caption{Accuracy}
        \end{subfigure}
        \vskip\baselineskip
        \begin{subfigure}[b]{.49\columnwidth}
             \includegraphics[width=.95\columnwidth]{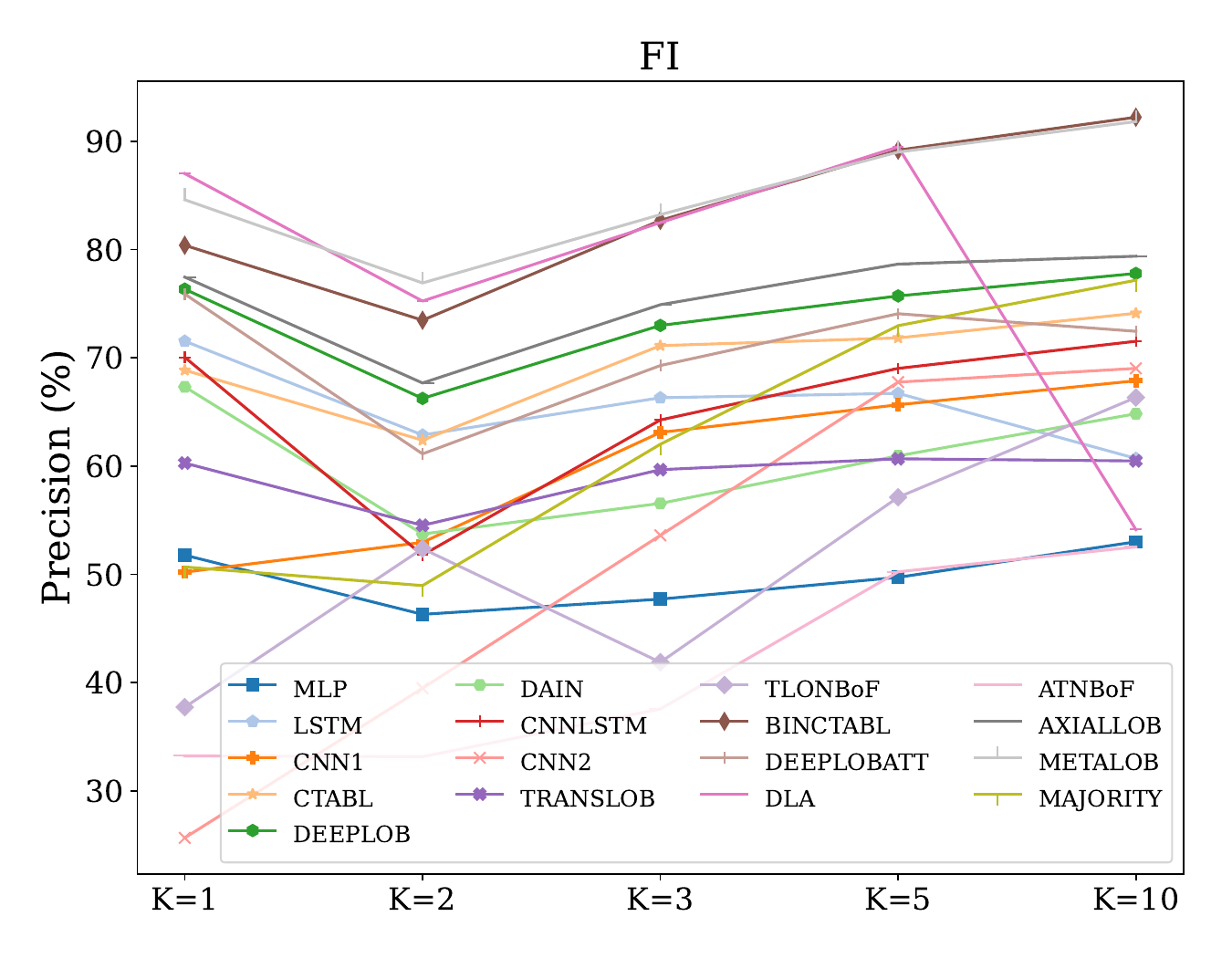}
            \caption{Precision}
            \label{fig:precF12010}
        \end{subfigure}
        \hfill
        \begin{subfigure}[b]{.49\columnwidth}
            \includegraphics[width=.95\columnwidth]{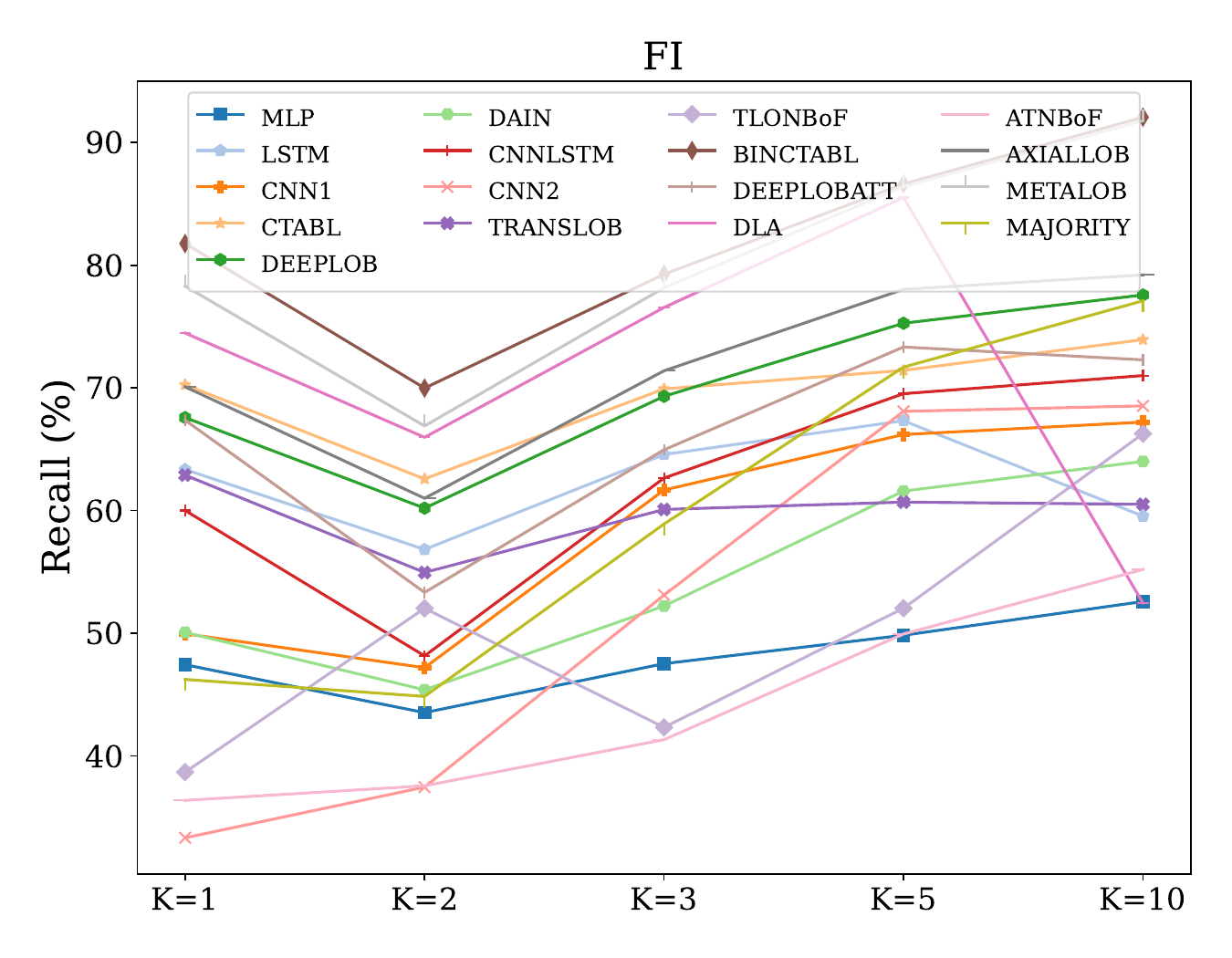}
            \caption{Recall}
            \label{fig:recallF12010}
        \end{subfigure}
       \caption{Evaluation metrics on different horizons $\mathcal{K}$ on FI-2010 dataset.}
        \label{fig:line_horiz_FI2010}
\end{figure}

\paragraph{Robustness} Figure~\ref{fig:line_horiz_FI2010} depicts the F1 score, accuracy, precision, and recall of the surveyed models obtained through our framework called LOBCAST, for the time horizons $\mathcal{K}= \{1,2,3,5,10\}$. 
Most of the models show similar behaviour with respect to the prediction horizons. In particular, regarding the F1-score, the worst performance is obtained for $k = 2$, after which there is an increasing trend as the prediction horizon increases. This might sound counterintuitive, as it consists of forecasting the price trend in a more distant future. However, for very short horizons, the labelling system adopted may be susceptible to noise affecting the model's capability to extract relevant patterns.

\begin{figure}
\centering
    \includegraphics[width=.7\linewidth]{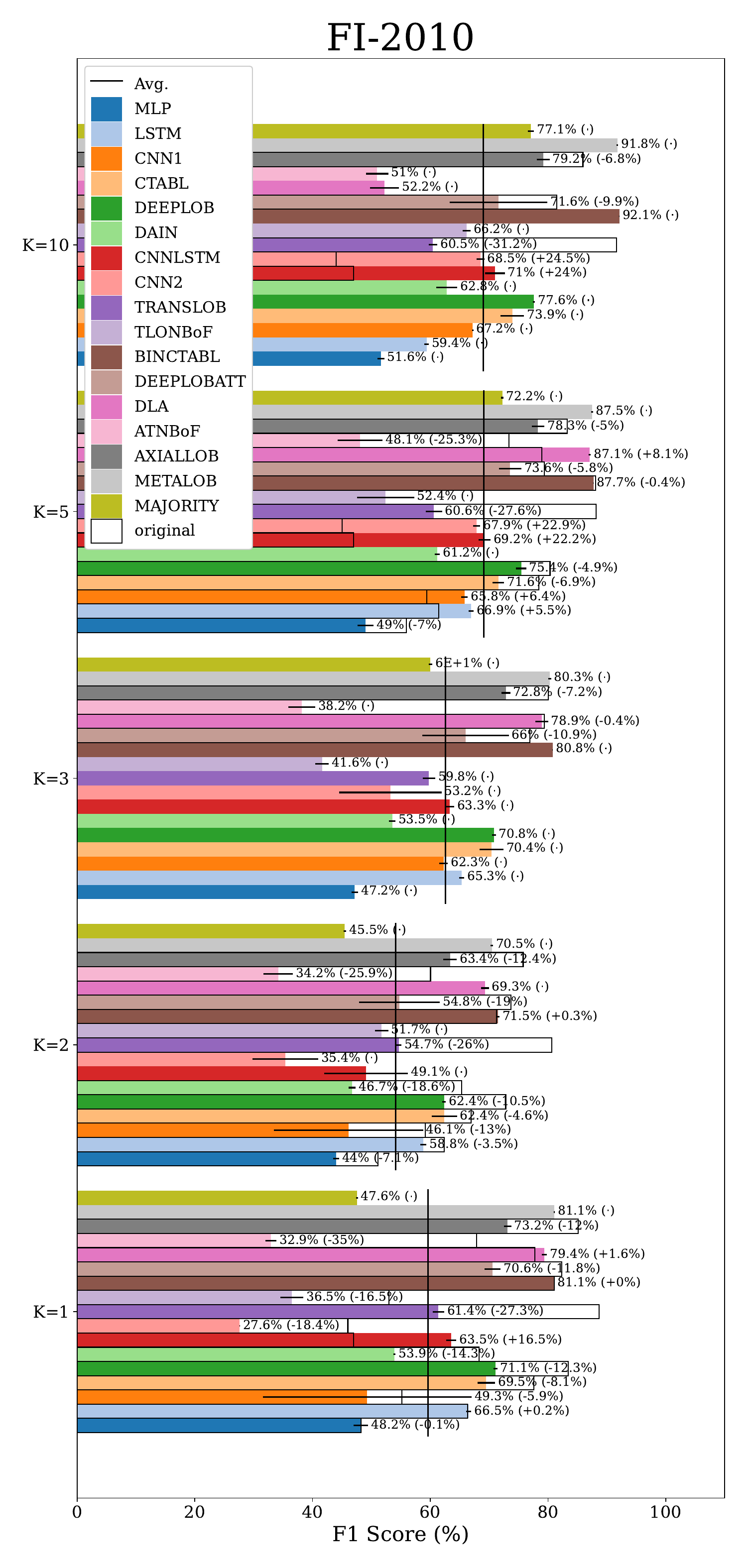}
    \caption{F1-score on FI-2010.}
        \label{fig:F1score_bar_FI2010}
\end{figure}

Figure~\ref{fig:F1score_bar_FI2010} shows a bar chart representing the F1-score of the 17 models reproduced using the LOBCAST framework for the five prediction horizons $\mathcal{K}$. 
The plot shows black empty bars representing the declared performance in the corresponding paper, when applicable. In the figure, the number on the bar represents the obtained performance in LOBCAST on the FI-2010 dataset, and the value in brackets indicates the difference between the obtained performance and the originally declared performance in the respective paper.
We highlight that not all papers declare their performance for all the horizons. 
The figure clearly highlights how robust the considered models are. Surprisingly, for CNN2 and CNNLSTM, our experiments achieved noticeably higher performance than the one declared in the original paper. We also observe the trend where the BINCTABL model consistently emerges as one of the top-performing models across all the horizons. Moreover, it notably closely aligns with the performance reported in the paper presenting it.

The largest discrepancy is observed for TRANSLOB and ATNBoF (or TNBoF-TA), whose average performances differ by 28\% from the original results.
On average, ATNBoF achieves an F1-score of only $40.9\%$. This substantial deviation from the claimed performance highlights the challenges and limitations associated with this particular model. 

\begin{figure}
    \centering
    \includegraphics[width=1.15\columnwidth]{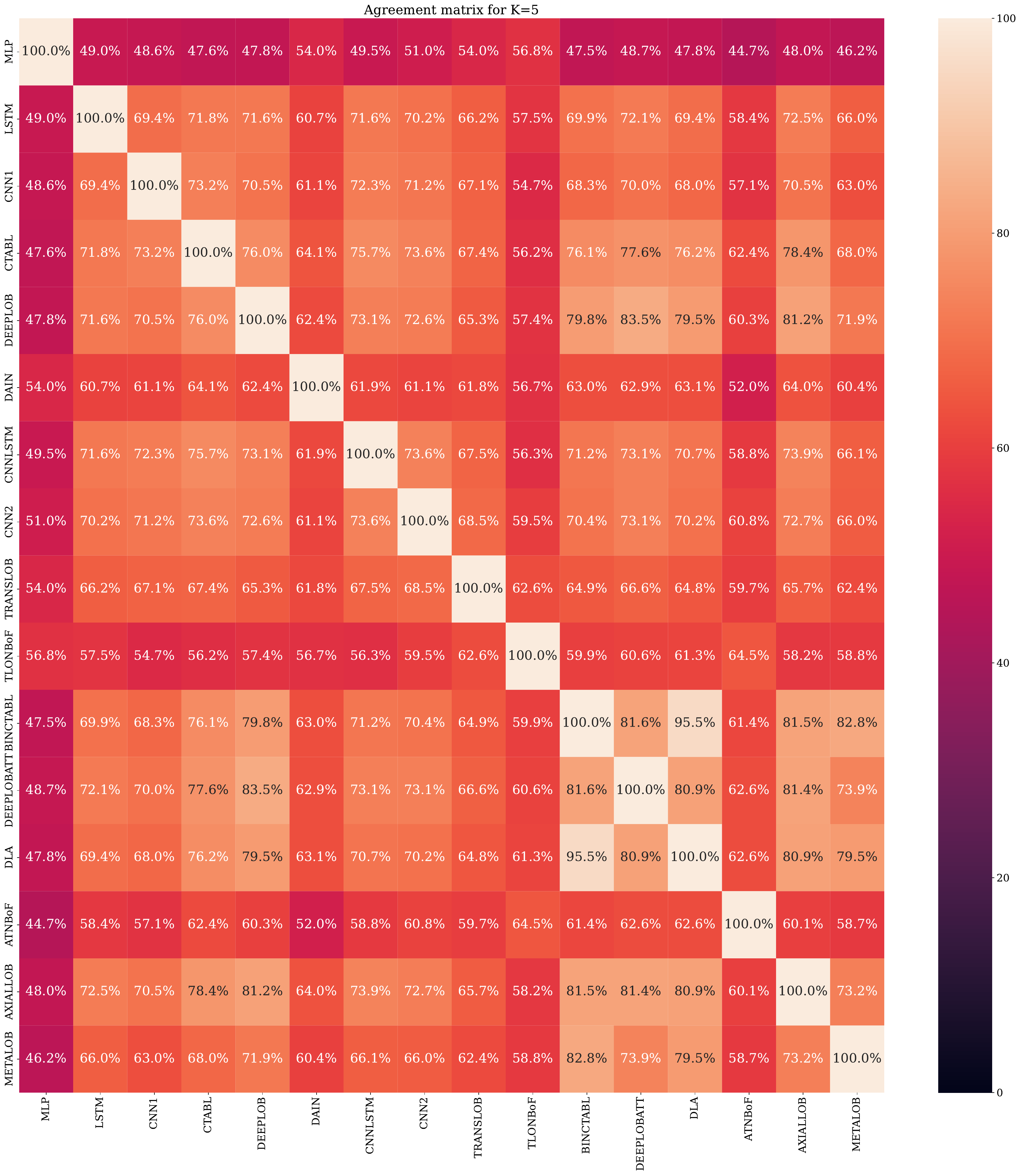}
    \caption{Agreement matrix FI-2010 in the horizon $k=5$.}
    \label{fig:agreementFI2010} 
\end{figure}

Figure~\ref{fig:agreementFI2010} shows the agreement matrix of the models for the horizon $k=5$. As expected, the highest agreement ($\approx$80\%) is among the best-performing models, namely BINCTABL, AXIALLOB, DEEPLOB, CTABL, DLA and DEEPLOBATT. The model that exhibits less correlation with the other models is MLP. 

The best-performing model in our benchmark is BINCTABL reaching 92.1\% of F1-Score on time horizon $k=10$. Specifically, BINCTABL introduces an Adaptive Bilinear Normalization layer to CTABL, enabling joint normalization of the input time series along both temporal and feature dimensions. This enhancement yields a remarkable improvement, with an average increase of $9.2\%$ in the F1-score compared to the second-best model (DLA).
Interestingly BINCTABL is composed only of 11.446 parameters, which makes it very fast at inference time (0.0005s).

\begin{figure}[h!]
        \centering
        \begin{subfigure}[b]{.49\columnwidth}
            \includegraphics[width=.95\columnwidth]{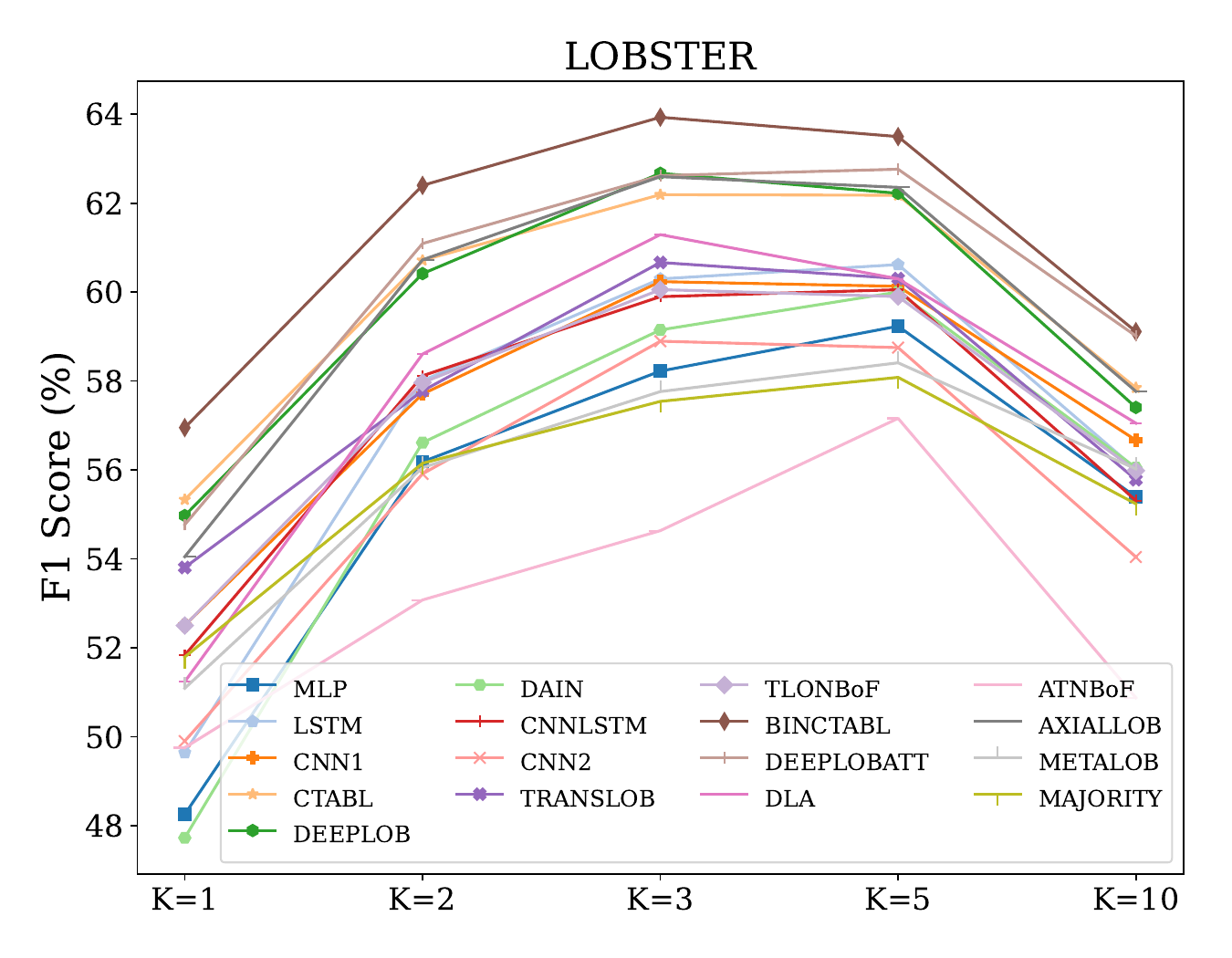}
            \caption{F1-Score}
        \end{subfigure}
        \begin{subfigure}[b]{.49\columnwidth}
            \includegraphics[width=.95\columnwidth]{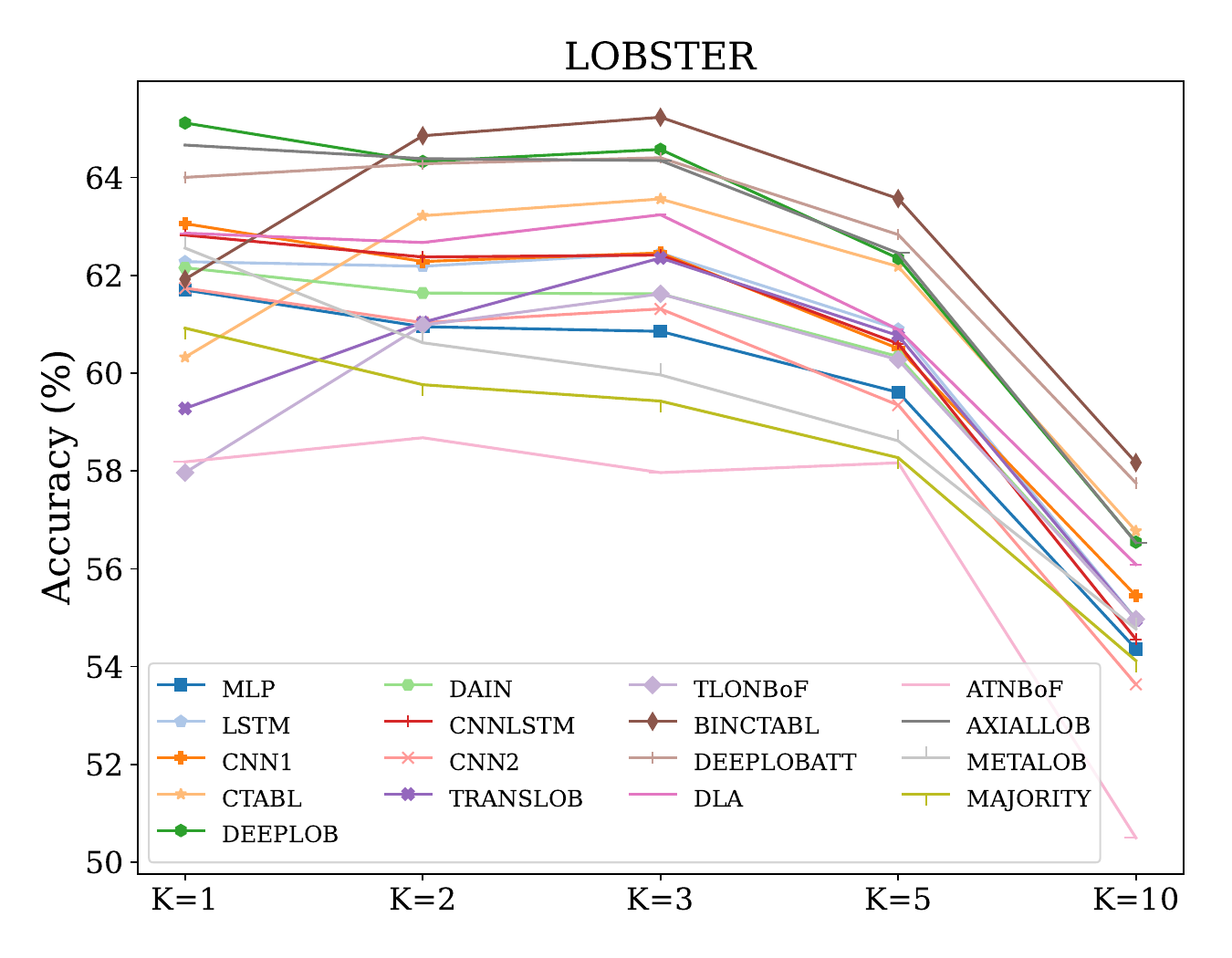}
            \caption{Accuracy}
        \end{subfigure}
        \vskip\baselineskip
        \begin{subfigure}[b]{.49\columnwidth}
             \includegraphics[width=.95\columnwidth]{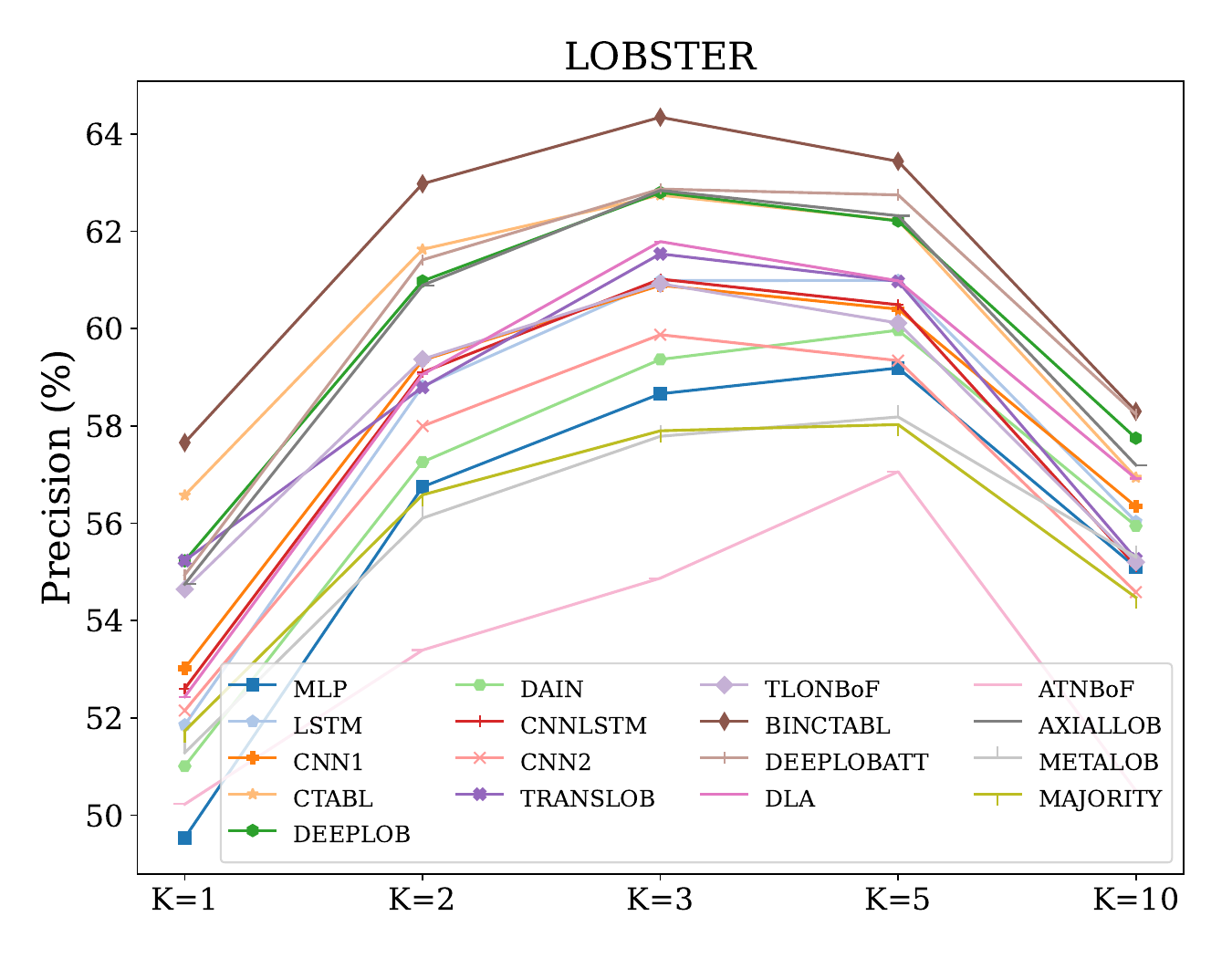}
            \caption{Precision}
        \end{subfigure}
        \hfill
        \begin{subfigure}[b]{.49\columnwidth}
            \includegraphics[width=.95\columnwidth]{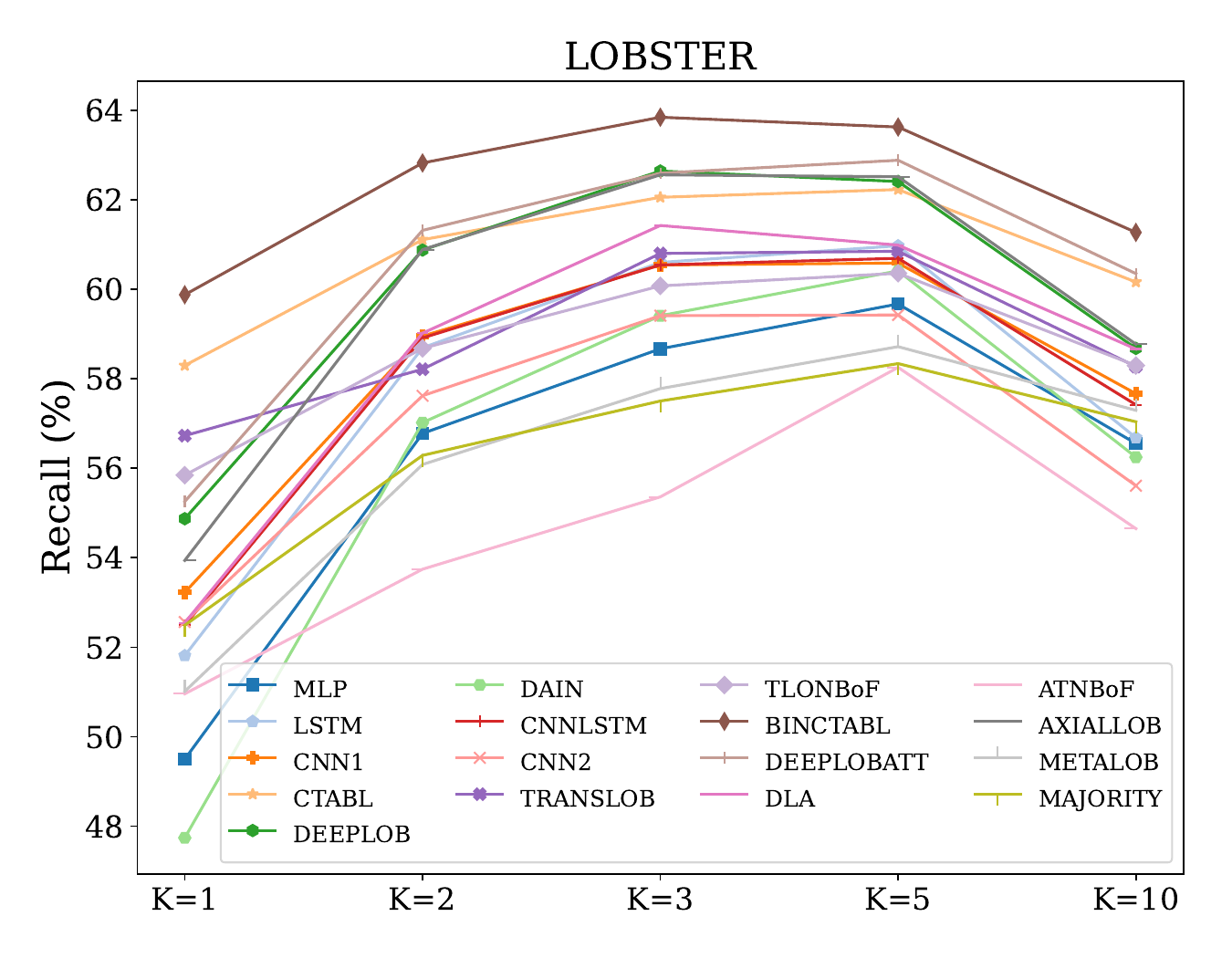}
            \caption{Recall}
        \end{subfigure}
       \caption{Evaluation metrics on different horizons $\mathcal{K}$ on LOB-2021.}
        \label{fig:line_horiz_LOB2021}
\end{figure}

\paragraph{Generalizability}
In this section, we provide additional results on the generalizability of the models. We evaluate the performance of the models on two different datasets: LOB-2021 and LOB-2022. The evaluation metrics used include F1-score, accuracy, precision, and recall, which are displayed in Figure~\ref{fig:line_horiz_LOB2021} for LOB-2021. The plot for LOB-2022 is omitted since they show similar properties.

We observe that most models exhibit a similar trend in both LOB-2021 and LOB-2022 datasets.
However, the performance curves in these generalizability tests differ from the results obtained on the FI-2010 dataset, shown in Figure~\ref{fig:line_horiz_FI2010}. 
Specifically, for the LOB-2021/2022 datasets, the F1-score of most models shows an increasing trend as the prediction horizon increases up to $k = 3$, after which it starts to decrease. 

To ease readability, in Table~\ref{tab:F1scoretab} we report the F1-score of all the models, horizons and periods.

\begin{figure}[t]

\resizebox{1\columnwidth}{!}{
    \centering
    \renewcommand{\arraystretch}{1.5} 
\begin{tabular}{|>{\centering\raggedleft}p{2.1cm}||>
     {\centering\arraybackslash}p{.6cm}|>
     {\centering\arraybackslash}p{.6cm}|>
     {\centering\arraybackslash}p{.6cm}|>
     {\centering\arraybackslash}p{.6cm}||>
     {\centering\arraybackslash}p{.6cm}|>
     {\centering\arraybackslash}p{.6cm}|>
     {\centering\arraybackslash}p{.6cm}|>
     {\centering\arraybackslash}p{.6cm}||>
     {\centering\arraybackslash}p{.6cm}|>
     {\centering\arraybackslash}p{.6cm}|>
     {\centering\arraybackslash}p{.6cm}|>
     {\centering\arraybackslash}p{.6cm}||>
     {\centering\arraybackslash}p{.6cm}|>
     {\centering\arraybackslash}p{.6cm}|>
     {\centering\arraybackslash}p{.6cm}|>
     {\centering\arraybackslash}p{.6cm}||>
     {\centering\arraybackslash}p{.6cm}|>
     {\centering\arraybackslash}p{.6cm}|>
     {\centering\arraybackslash}p{.6cm}|>
     {\centering\arraybackslash}p{.6cm}||
     } 
    \hline 
& \multicolumn{4}{c||}{$k = 1$} & \multicolumn{4}{c||}{$k = 2$} & \multicolumn{4}{c||}{$k = 3$} & \multicolumn{4}{c||}{$k = 5$} & \multicolumn{4}{c||}{$k = 10$} \\ \hline
\textbf{Model} & \textbf{FI 2010} &  \textbf{FI$^r$ 2010} & \textbf{LOB 2021} &  \textbf{LOB 2022} &  \textbf{FI 2010} &  \textbf{FI$^r$ 2010} &  \textbf{LOB 2021} &  \textbf{LOB 2022} & \textbf{FI 2010} &  \textbf{FI$^r$ 2010} & \textbf{LOB 2021} &  \textbf{LOB 2022} & \textbf{FI 2010} &  \textbf{FI$^r$ 2010} & \textbf{LOB 2021} &  \textbf{LOB 2022} & \textbf{FI 2010} &  \textbf{FI$^r$ 2010} & \textbf{LOB 2021} &  \textbf{LOB 2022} \\ \hline\hline
MLP & $48.3$ & $48.2$ & $48.3$ & $51.1$ & $51.1$ & $44.0$ & $56.2$ & $54.1$ & -- & $47.2$ & $58.2$ & $55.9$ & $56.0$ & $49.0$ & $59.2$ & $55.0$ & -- & $51.6$ & $55.4$ & $49.3$ \\ \hline
LSTM & $66.3$ & $66.5$ & $49.6$ & $53.7$ & $62.4$ & $58.8$ & $58.0$ & $57.4$ & -- & $65.3$ & $60.3$ & $60.6$ & $61.4$ & $66.9$ & $60.6$ & $56.2$ & -- & $59.4$ & $56.0$ & $52.6$ \\ \hline
CNN1 & $55.2$ & $49.3$ & $52.5$ & $55.3$ & $59.2$ & $46.1$ & $57.7$ & $59.8$ & -- & $62.3$ & $60.2$ & $59.3$ & $59.4$ & $65.8$ & $60.1$ & $58.5$ & -- & $67.2$ & $56.7$ & $52.6$ \\ \hline
CTABL & $77.6$ & $69.5$ & $55.3$ & $57.8$ & $66.9$ & $62.4$ & $60.7$ & $60.9$ & -- & $70.4$ & $62.2$ & $60.8$ & $78.4$ & $71.6$ & $62.2$ & $58.8$ & -- & $73.9$ & $57.8$ & $52.0$ \\ \hline
DEEPLOB & $83.4$ & $71.1$ & $55.0$ & $57.0$ & $72.8$ & $62.4$ & $60.4$ & $\mathbf{62.0}$ & -- & $70.8$ & $62.7$ & $\mathbf{62.4}$ & $80.4$ & $75.4$ & $62.2$ & $60.8$ & -- & $77.6$ & $57.4$ & $55.2$ \\ \hline
DAIN & $68.3$ & $53.9$ & $47.7$ & $52.2$ & $65.3$ & $46.7$ & $56.6$ & $54.9$ & -- & $53.5$ & $59.1$ & $55.8$ & -- & $61.2$ & $60.0$ & $56.5$ & -- & $62.8$ & $56.1$ & $51.2$ \\ \hline
CNNLSTM & $47.0$ & $63.5$ & $51.8$ & $55.0$ & -- & $49.1$ & $58.1$ & $59.8$ & -- & $63.3$ & $59.9$ & $59.2$ & $47.0$ & $69.2$ & $60.1$ & $57.1$ & $47.0$ & $71.0$ & $55.3$ & $53.1$ \\ \hline
CNN2 & $46.0$ & $27.6$ & $49.9$ & $51.9$ & -- & $35.4$ & $55.9$ & $59.0$ & -- & $53.2$ & $58.9$ & $58.7$ & $45.0$ & $67.9$ & $58.8$ & $57.3$ & $44.0$ & $68.5$ & $54.0$ & $52.0$ \\ \hline
TRANSLOB & $\mathbf{88.7}$ & $61.4$ & $53.8$ & $43.7$ & $\mathbf{80.6}$ & $54.7$ & $57.8$ & $43.0$ & -- & $59.8$ & $60.7$ & $57.5$ & $\mathbf{88.2}$ & $60.6$ & $60.3$ & $56.6$ & $\mathbf{91.6}$ & $60.5$ & $55.8$ & $51.0$ \\ \hline
TLONBoF & $53.0$ & $36.5$ & $52.5$ & $53.1$ & -- & $51.7$ & $58.0$ & $56.5$ & -- & $41.6$ & $60.1$ & $57.1$ & -- & $52.4$ & $59.9$ & $55.7$ & -- & $66.2$ & $56.0$ & $48.5$ \\ \hline
BINCTABL & $81.0$ & $\mathbf{81.1}$ & $\mathbf{57.0}$ & $\mathbf{58.4}$ & $71.2$ & $\mathbf{71.5}$ & $\mathbf{62.4}$ & $\mathbf{62.0}$ & -- & $\mathbf{80.8}$ & $\mathbf{63.9}$ & $62.2$ & $88.1$ & $\mathbf{87.7}$ & $\mathbf{63.5}$ & $60.4$ & -- & $\mathbf{92.1}$ & $\mathbf{59.1}$ & $53.2$ \\ \hline
DEEPLOBATT & $82.4$ & $70.6$ & $54.8$ & $55.8$ & $73.7$ & $54.8$ & $61.1$ & $60.5$ & $76.9$ & $66.0$ & $62.6$ & $62.1$ & $79.4$ & $73.6$ & $62.8$ & $\mathbf{60.9}$ & $81.5$ & $71.6$ & $59.0$ & $\mathbf{55.3}$ \\ \hline
DLA & $77.8$ & $79.4$ & $51.2$ & $54.4$ & -- & $69.3$ & $58.6$ & $58.0$ & $79.4$ & $78.9$ & $61.3$ & $60.0$ & $79.0$ & $87.1$ & $60.3$ & $57.3$ & -- & $52.2$ & $57.1$ & $53.4$ \\ \hline
ATNBoF & $67.9$ & $32.9$ & $49.8$ & $47.8$ & $60.0$ & $34.2$ & $53.1$ & $50.3$ & -- & $38.2$ & $54.6$ & $41.3$ & $73.4$ & $48.1$ & $57.2$ & $59.8$ & -- & $51.0$ & $50.9$ & $40.9$ \\ \hline
AXIALLOB & $85.1$ & $73.2$ & $54.0$ & $56.9$ & $75.8$ & $63.4$ & $60.7$ & $60.1$ & $\mathbf{80.1}$ & $72.8$ & $62.6$ & $62.0$ & $83.3$ & $78.3$ & $62.4$ & $59.6$ & $85.9$ & $79.2$ & $57.8$ & $54.6$ \\ \hline\hline
METALOB & -- & $81.1$ & $51.1$ & $52.3$ & -- & $70.5$ & $56.1$ & $53.3$ & -- & $80.3$ & $57.8$ & $55.3$ & -- & $87.5$ & $58.4$ & $54.5$ & -- & $91.8$ & $56.0$ & $50.9$ \\ \hline
MAJORITY & -- & $47.1$ & $51.8$ & $50.6$ & -- & $44.9$ & $56.2$ & $49.2$ & -- & $59.7$ & $57.5$ & $48.1$ & -- & $71.8$ & $56.9$ & $46.7$ & -- & $76.3$ & $55.2$ & $44.7$ \\ \hline
\end{tabular}
}
\captionof{table}{F1-score on LOB-2021 and LOB-2022. Columns FI 2010, FI$^r$ 2010, LOB 2021, and LOB 2022, respectively represent the claimed performance of the models in the respective papers, the performance reproduced with LOBCAST on FI, LOB-2021, LOB-2022.}

\label{tab:F1scoretab}
\end{figure}

The performance of the models, as reported by the authors of the selected paper, exhibits changes when evaluated on the LOB-2021 and LOB-2022 datasets. These changes show varying degrees of generalizability among the models.

Notably, the ATNBoF model demonstrates the most substantial improvement with respect to the declared performances, showing an average increase of 12.2\% across all prediction horizons. A similar improvement is exhibited by MLP and TLONBoF.
Despite this improvement, ATNBoF still exhibits the lowest overall performance with an average score of 53.1\%. It is worth mentioning that ATNBoF is the most sensitive to random initialization.

In contrast, the other models experience a significant decline in performance when evaluated on LOB-2021 and LOB-2022 datasets. For example, the previously best-performing model on the FI-2010 dataset, BINCTABL, shows an average decrease in F1-score of approximately 19.6\% across all prediction horizons. This decline results in a generalizability score of 73.5\% (as mentioned in Table 2 of the main paper). However, despite this decline, BINCTABL remains the top-performing model when evaluated on the LOB-2021 dataset on almost all the prediction horizons. On these datasets it exhibits similar performance to DEPPLOB and DEEPLOBATT models.

\begin{figure}
\centering
        \includegraphics[width=1.15\columnwidth]{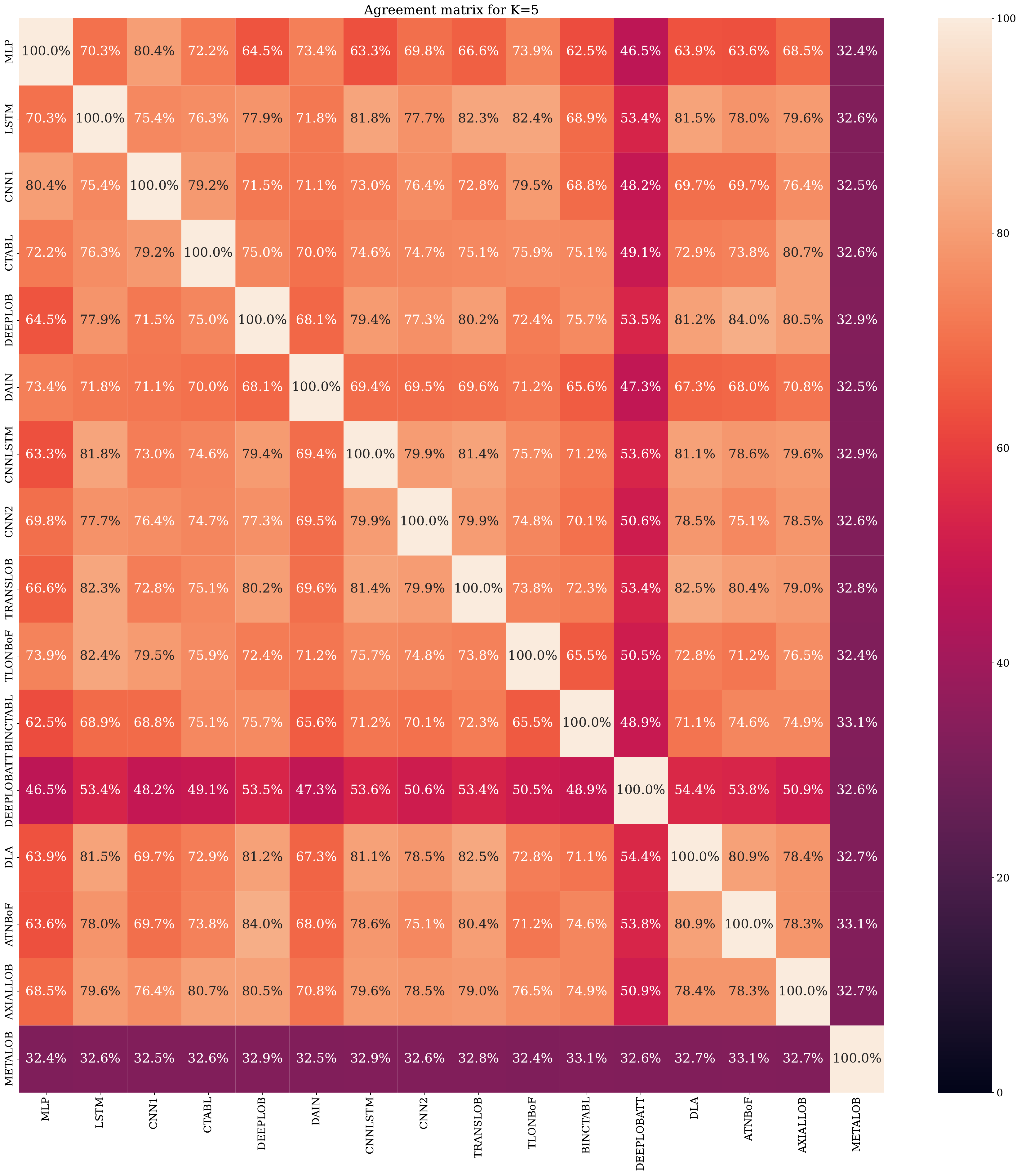}
        \caption{Agreement matrix on LOB-2022.}
        \label{fig:agreeement2022}
\end{figure}

Figure~\ref{fig:agreeement2022} shows the agreement matrix on LOB-2022.  
Considering the more flattened performances of the models on the LOB-2021/2022 dataset compared to the FI-2010 dataset, the agreement percentages among the models are consistently high, and no distinct patterns are observed. Unlike FI-2010, where METALOB predicted as BINCTABL 82.8\% of the time, on LOB-2022 (and also LOB-2021) METALOB showed no preference for any model, resulting in a balanced agreement rate ($\approx33\%$) among all models.
We decided not to include the agreement matrix of LOB-2021 because it was similar to LOB-2022.

\begin{figure}
   \centering
\includegraphics[width=\columnwidth]{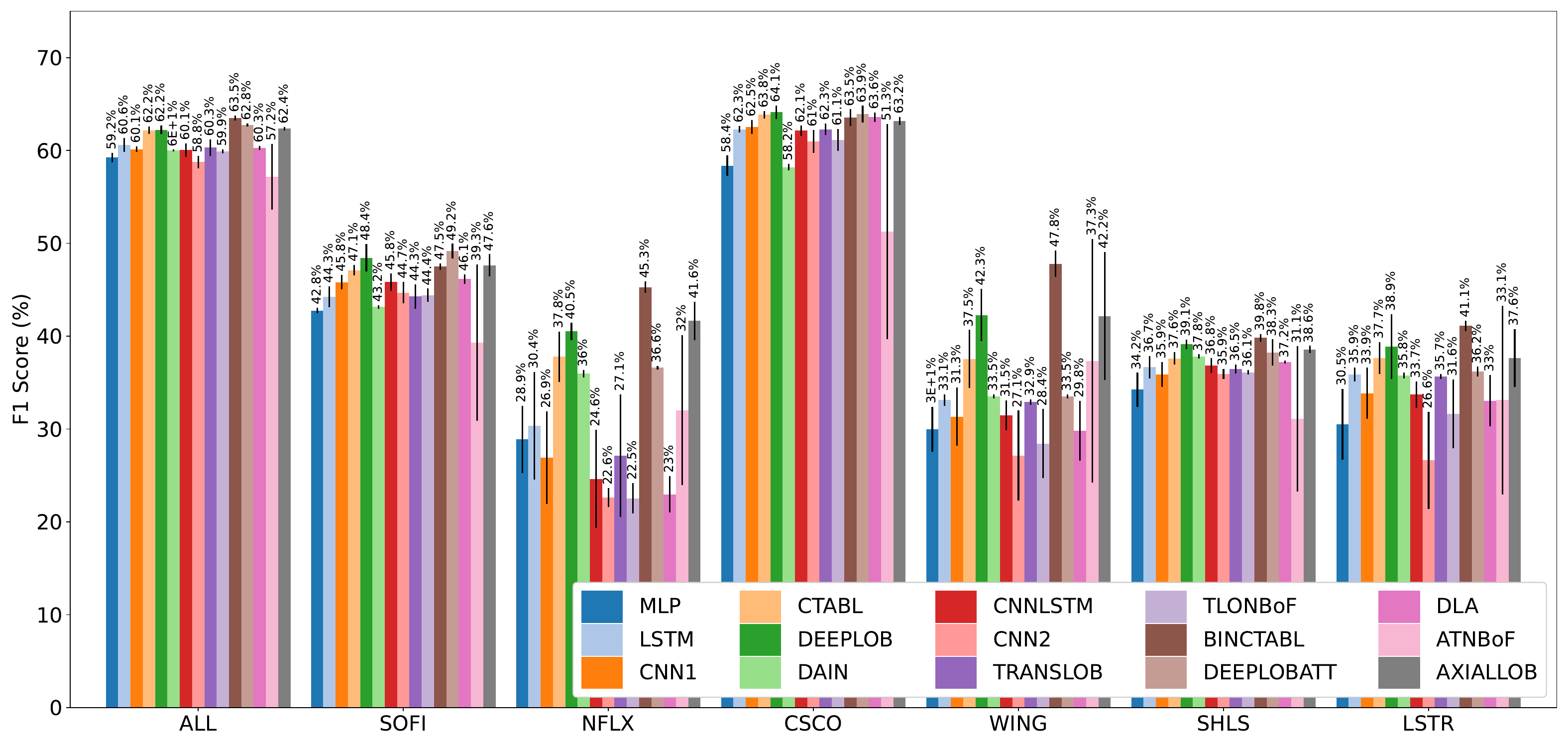}  
    \caption{F1-Score per stock, time horizon $k=5$, on LOB-2021.}
    \label{fig:per_stock}
\end{figure}

In Figure~\ref{fig:per_stock}, we present the results of our tests for the time horizon $k=5$ on each individual stock from LOB-2021 dataset.
Among the tested stocks, CSCO stands out as yielding the highest performance. This may be attributed to the high
stationarity of CSCO (balance 18-65-17\% in the train set), indicating more stable and predictable
behaviour. This hypothesis is supported by the confusion matrices reported in the main paper in section 4.4, which shows the best performance in the stationary class for BINCTABL model.
We highlight that it was impossible to extract the per-stock information on the FI-2010 dataset because it was already assembled, and the authors did not provide information on that procedure.

\paragraph{Labelling}

\begin{figure}
    \begin{subfigure}[b]{\textwidth}
        \centering
        \includegraphics[width=.85\columnwidth]{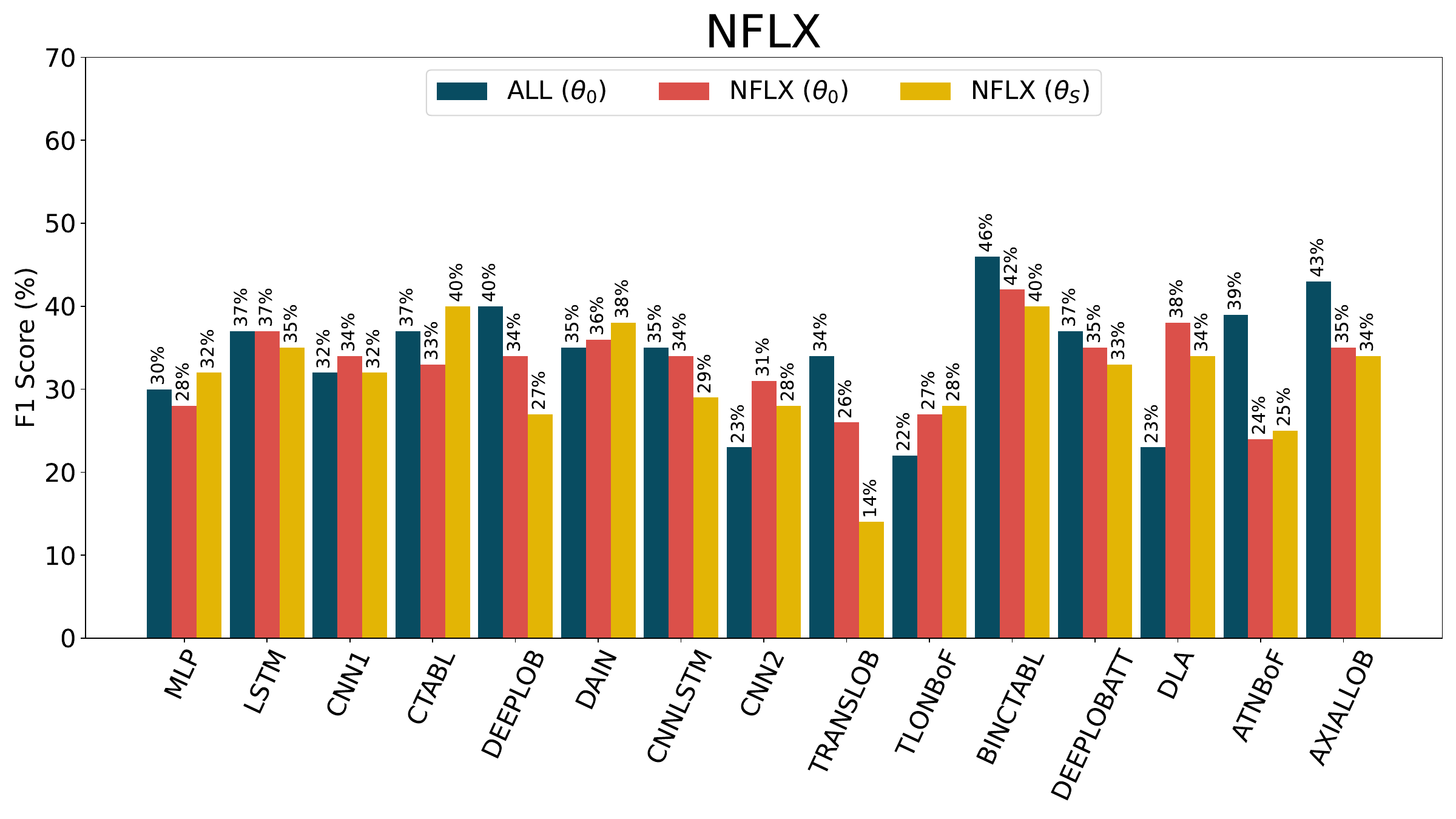}
        \vspace{-0.1in}
        \caption{NFLX}
        \label{fig:alfa_net}
    \end{subfigure}
    \begin{subfigure}[b]{\textwidth}
        \centering
        \includegraphics[width=.85\columnwidth]{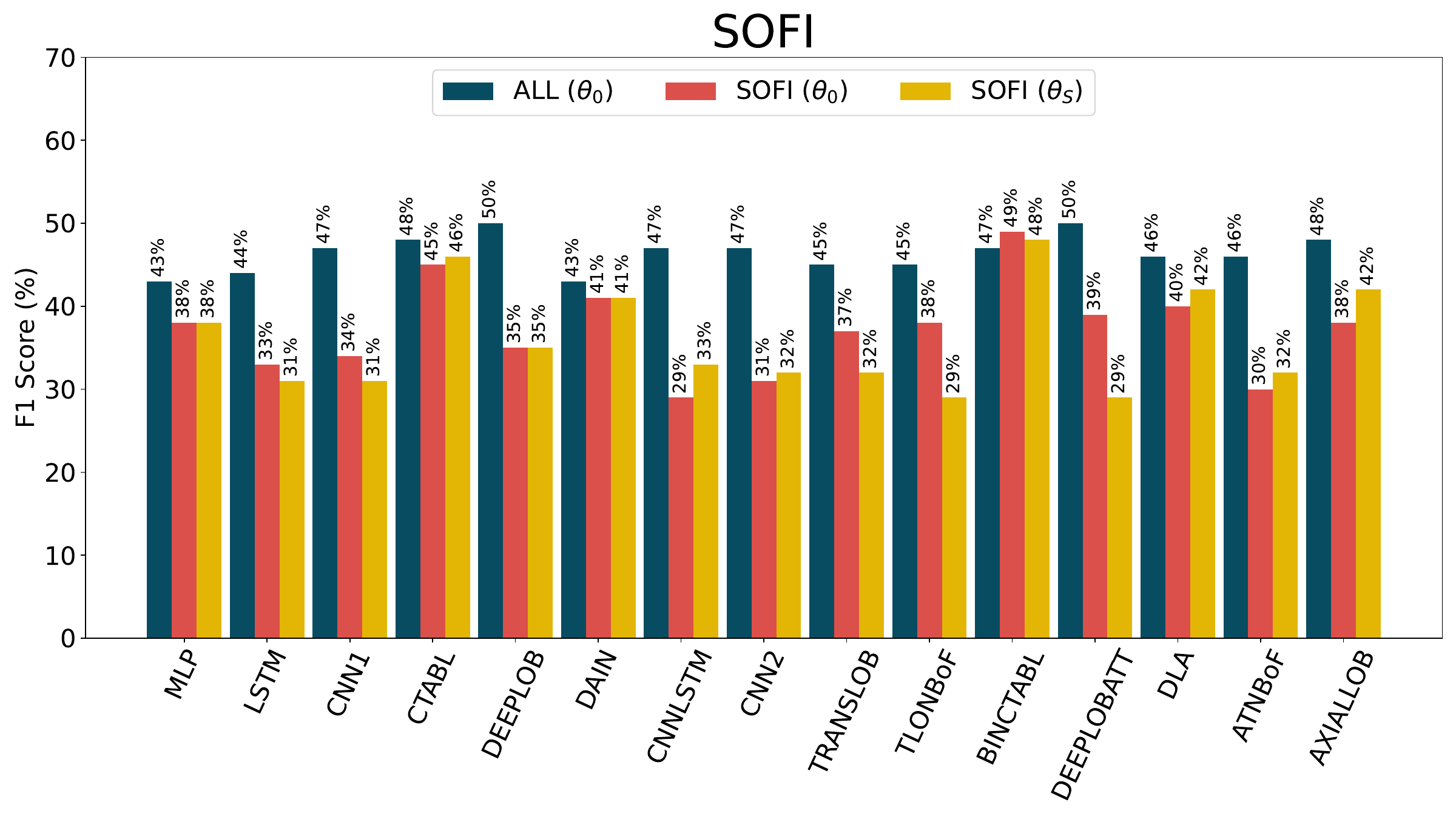}
        \vspace{-0.1in}
        \caption{SOFI}
        \label{fig:alfa_sof}
    \end{subfigure}
    \caption{Different labelling strategies on NFLX and SOFI stocks for $k=5$.}
    \label{fig:alphatest}
\end{figure}

The experiments shown above highlight that the models' performance does not exhibit a clear trend with respect to the prediction horizon. 
The labelling method is probably the cause of this phenomenon, in fact, classifying trends based on the mid-price tends to embody noise on the nearer horizons. This hypothesis is supported by the work of Zhang et al. \cite{zhang2019deeplob}, specifically, they generated a dataset using an alternative labelling method  that relies on the mean of the previous and next $k$ mid-prices to identify trends. Interestingly they observed an inverse trend in performance with respect to the horizons; in fact, the best performances were achieved with the shortest horizon and deteriorated as it increased. 
While exploring various labelling techniques is beyond the scope of this benchmark, we provide an initial investigation in this direction. Specifically, focusing on $k = 5$ in LOB-21, we select two stocks, NFLX and SOFI.

Based on Equation 1 and 2 of the original paper, we can define $\theta_N$ and $\theta_S$ as the thresholds that balance the occurrences of the classes for the stocks NFLX and SOFI, respectively. Similarly, we can define $\theta_0$ as the threshold that balances the occurrences of the classes for the ensemble of six stocks within the dataset.

Figure~\ref{fig:alphatest} shows the results of three different training settings: (i.) \textbf{ALL (}$\mathbf{\theta_0}$\textbf{)} represents the training of the models over the ensemble of all the six stocks using the threshold $\theta_0$; (ii.) \textbf{NFLX (}$\mathbf{\theta}_0$\textbf{)} (\textbf{SOFI (}$\mathbf{\theta}_0$\textbf{)}) represents the training of the models over NFLX (SOFI) stock using the threshold $\theta_0$. (iii.) \textbf{NFLX (}$\mathbf{\theta}_N$\textbf{)} (\textbf{SOFI (}$\mathbf{\theta}_S$\textbf{)}) represents the training of the models over NFLX (SOFI) stock using the threshold $\theta_N$.

In the case of SOFI, all methods, except for BINCTABL, achieve the highest performance in the \textbf{ALL (}$\mathbf{\theta_0}$\textbf{)}  setting. This indicates that these models are able to extract useful signals from other stocks, reducing overfitting and improving overall performance.
On the other hand, comparing the \textbf{SOFI (}$\mathbf{\theta}_0$\textbf{)} and \textbf{SOFI (}$\mathbf{\theta}_S$\textbf{)} settings does not provide significant insights. This suggests that the balancing of the three classes is not crucial for achieving higher performance. This is even more the case for NFLX in Figure \ref{fig:alfa_net}, considering that the imbalance due to $\mathbf{\theta}_0$ is much higher (see Table \ref{tab:stocks}).

These results indicate that the labelling mechanism should be revised from its current definition and be agnostic with respect to the balancing involved. 
Trends definitions should not solely depend on the magnitude of the future price shift relative to the current price. Other factors, such as persistence over time and volume considerations, should also be taken into account. 
A more comprehensive discussion of the limitations and challenges associated with the labelling mechanisms can be found in the main paper, particularly in the final discussion and conclusions section. 

\section{Profit Analysis}

\begin{figure}
    \centering
    \begin{minipage}{.48\linewidth} 
    \includegraphics[width=\linewidth]{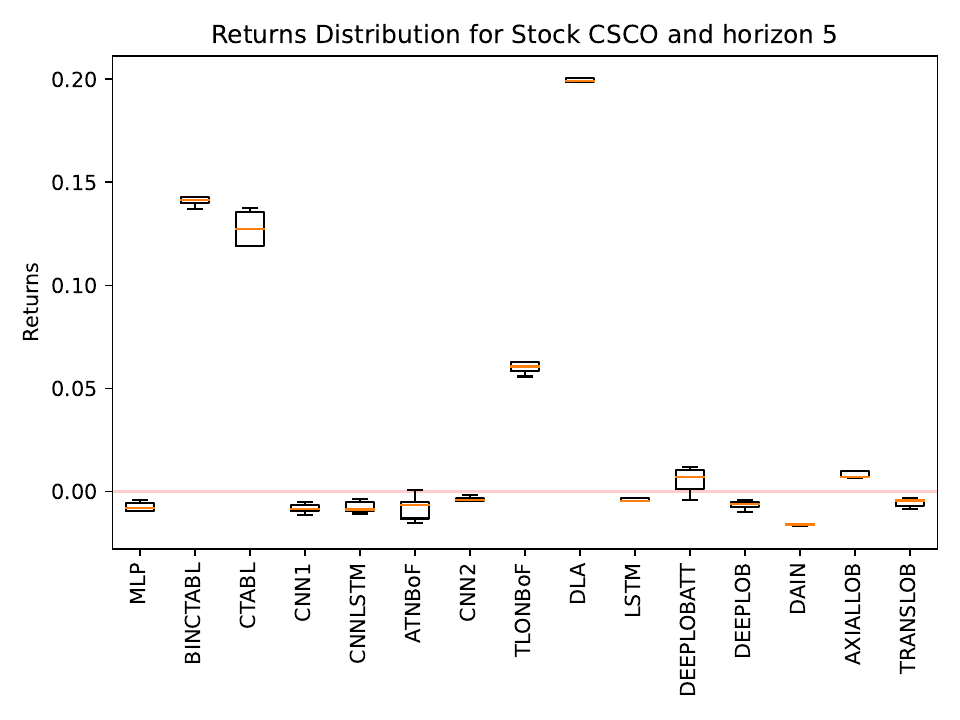}
    \label{fig:res_back_CSCO_5}
    \end{minipage}
    \hfill
    \begin{minipage}{.48\linewidth} 
    \includegraphics[width=\linewidth]{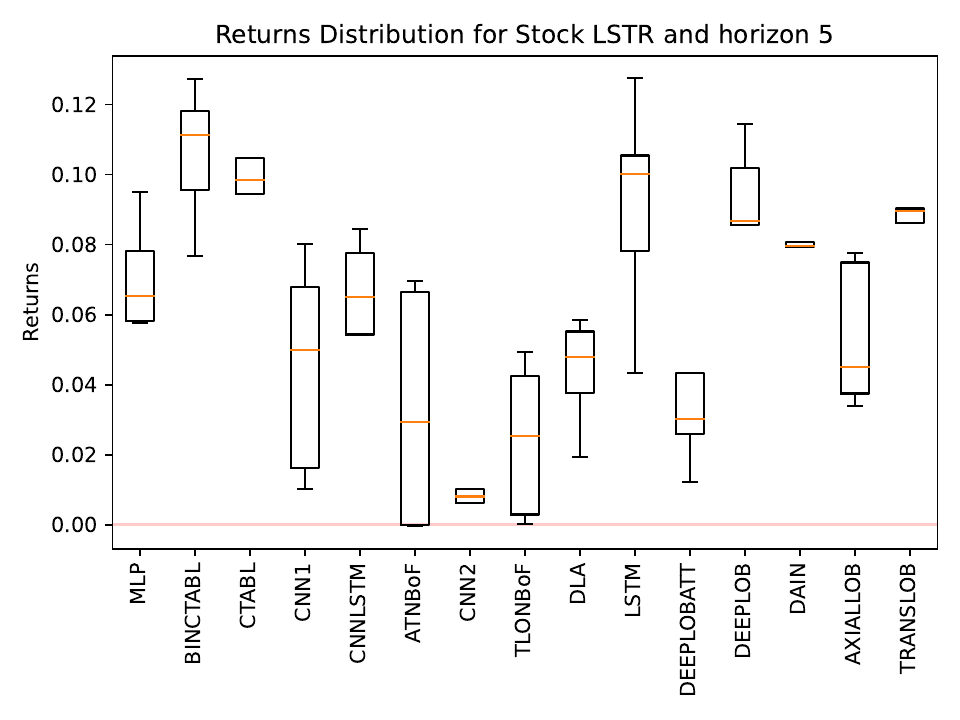}
    \label{fig:res_back_LSTR_5}
    \end{minipage}   
    \hfill
    \begin{minipage}{.48\linewidth} 
    \includegraphics[width=\linewidth]{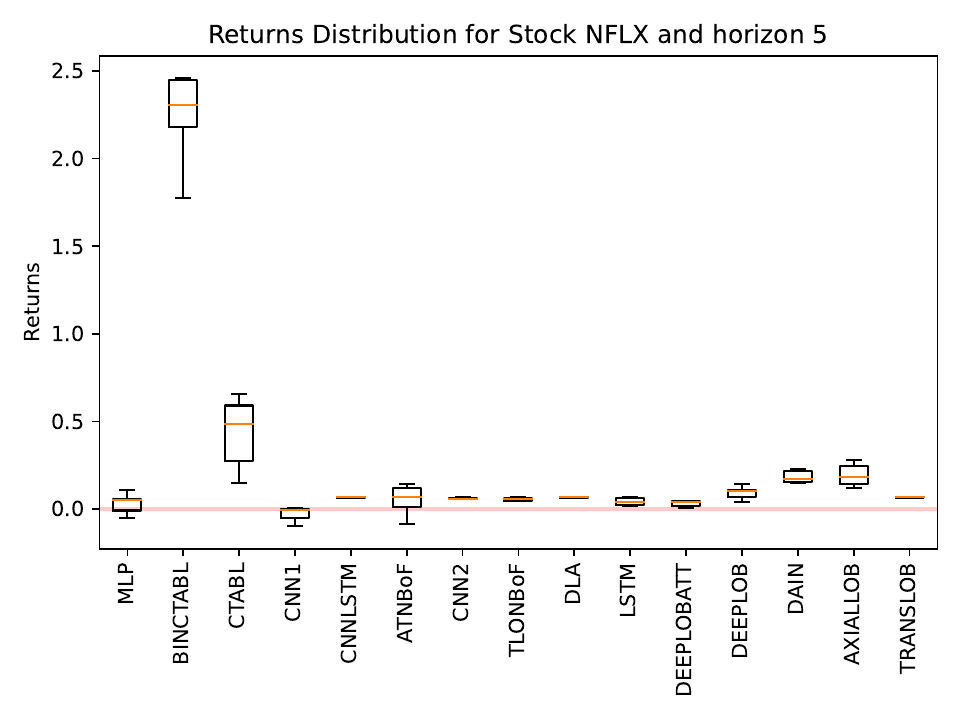}
    \label{fig:res_back_NFLX_5}
    \end{minipage}
    \hfill
    \begin{minipage}{.48\linewidth} 
    \includegraphics[width=\linewidth]{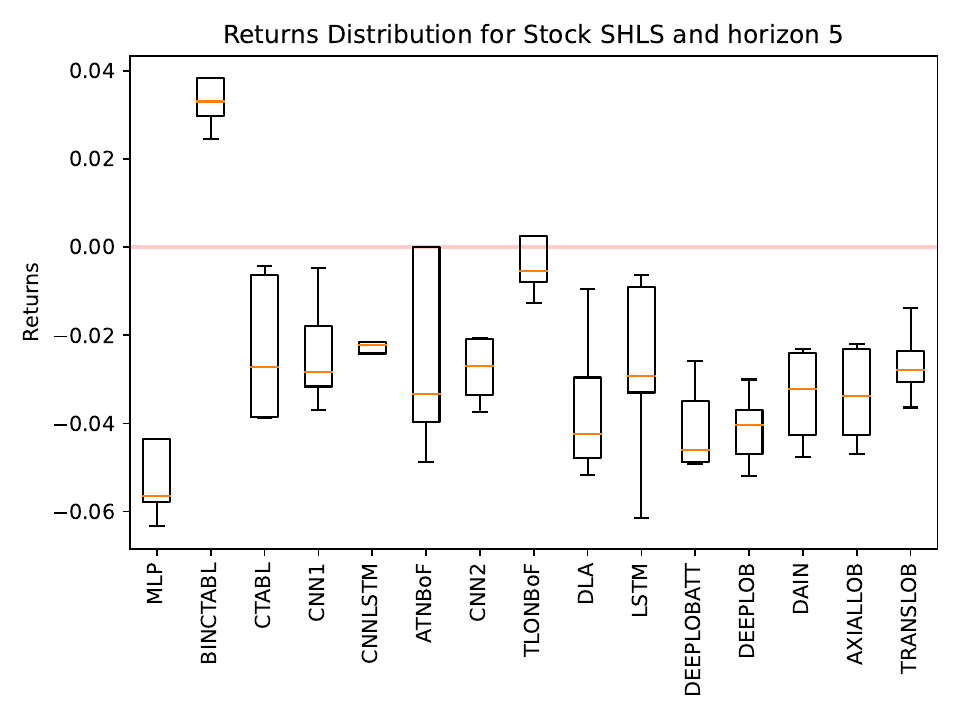}
    \label{res_back_SHLS_5}
    \end{minipage}
    \hfill
    \begin{minipage}{.48\linewidth} 
    \includegraphics[width=\linewidth]{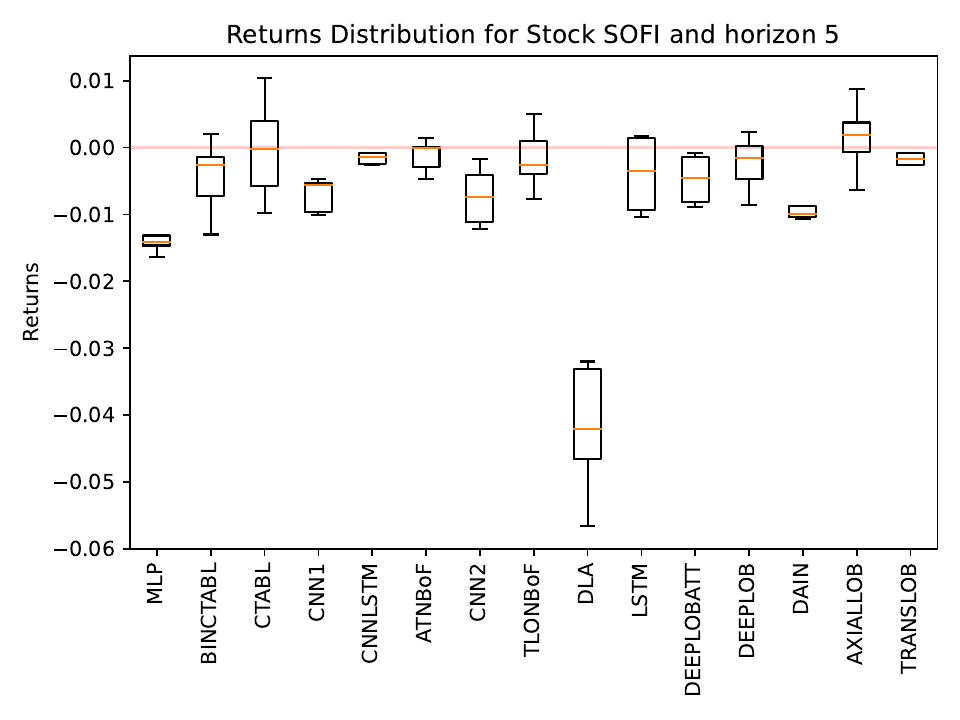}
    \label{fig:res_back_SOFI_5}
    \end{minipage}
    \hfill
    \begin{minipage}{.48\linewidth} 
    \includegraphics[width=\linewidth]{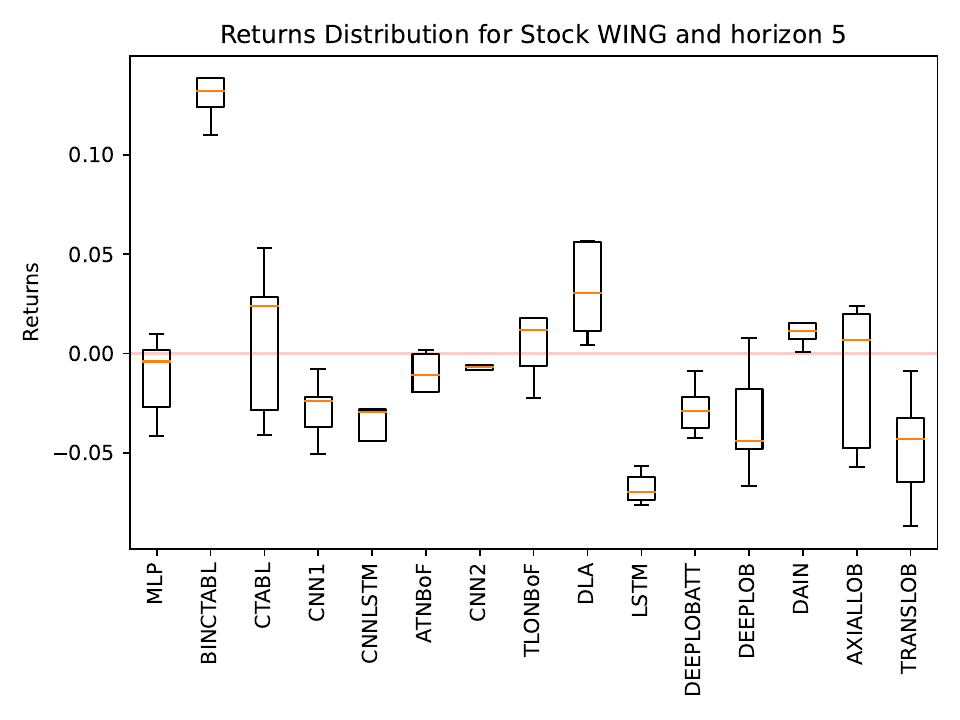}
    \label{fig:res_back_WING_5}
    \end{minipage}
\vspace{-3mm}
\caption{Distribution of returns on five seeds.}
\label{fig:returns}
\end{figure} 

As a final benchmark test, we conducted a trading simulation using our framework, relying on Backtesting.py Python library \footnote{https://kernc.github.io/backtesting.py/}.
As highlighted by \cite{olorunnimbe2023deep},  most of the existing literature in the \gls*{sptp} field neglects backtesting, even though it is essential for evaluating the performance of algorithmic trading strategies and for potential real-world use.

We performed backtesting using the same period as the test set of the LOB-2021 dataset, i.e., from 2021-07-13 to 2021-07-15. 
To perform backtesting, we generated an \gls*{ohlc} time series with a 10 events period. 
The \gls*{ohlc} is an aggregation technique to summarize periods of a time series, e.g., minutes, hours, days, or a number of events (10 in this case). 
Each data point of the series represents four aggregates of the considered period. The \textit{Open} represents the first price of the period; \textit{High} is the highest price of the period; \textit{Low} is the lowest price of the period; \textit{Close} is the last price of the period.

We base our trading simulation on the methodology of the seminal paper \cite{zhang2019deeplob} in this field, in which the authors conducted a similar experiment. 
We established certain parameters for our simulation. Firstly, we set the number of shares per trade to a fixed value of 1, simplifying our analysis and assuming a negligible market impact. Furthermore, our simulated trader begins with an initial capital of \$10.000, and we make the assumption of no transaction fees.

The \textit{trading strategy} relies on the models and operates by generating signals every 10 events to predict subsequent price movements. These signals, categorized as \textit{up}, \textit{stationary}, or \textit{down}, determine the trading action. When the signal is \textit{up}, the simulated trader places a buy order. Conversely, if the signal is \textit{down} and the trader currently holds a long position, he places a sell order. In cases where the signal is \textit{stationary}, the trader takes no action. The orders are filled at the next open price.

The results of the trading simulation for each stock are presented in Figure~\ref{fig:returns}.
The strongest correlation observed is between the daily returns of the stocks, as shown in Table~\ref{tab:stocks}, and the returns of the strategy described above. In fact, the two stocks with the highest positive daily returns (namely LSTR and NFLX) are the only ones for which the strategy is profitable. On the other hand, the two stocks with the highest negative daily returns (SOFI and SHLS) are the ones for which most models show a negative return. 
Another correlation, albeit less strong, is between the volatility of the stocks and the return of the models. Specifically, lower volatility is associated with higher model returns.

We recognize the limitations of this simulation. For instance, we do not perform portfolio optimization or position sizing, we assume the trades execution at the mid-price, and we ignore transaction costs, but a realistic and sophisticated algorithmic trading simulation is beyond the scope of this study and remains an interesting aspect for future research.

\bibliographystyle{IEEEtran}
\bibliography{IEEEabrv, main}

\end{document}